\documentclass[journal]{IEEEtran}
 
\usepackage{mathbf-abbrevs}
\newcommand{\reals}{{\mathbb R}}
\newcommand{\expect}{{\mathbb E}}

\newcommand{\ints}{{\mathbb Z}}

\newcommand{\covar}{\operatorname{cov}}
\newcommand{\prob}{{\operatorname{Pr}}}

\newcommand{\abs}[1]{{\left\vert #1 \right\vert}}
\newcommand{\sabs}[1]{{\vert #1 \vert}}

\newcommand{\round}[1]{{\left\lfloor #1 \right\rceil}}

\newcommand{\fracpart}[1]{\left\langle #1 \right\rangle}
\newcommand{\sfracpart}[1]{\langle #1 \rangle}

\usepackage[square,comma,numbers,sort&compress]{natbib}

\usepackage{units}
		
\usepackage{booktabs}
		
\usepackage{ifpdf}
\ifpdf
  \usepackage[pdftex]{graphicx}
\else
	\usepackage{graphicx}
\fi

\usepackage{amsmath,amsfonts,amssymb, amsthm, bm} 

\usepackage[vlined, linesnumbered]{algorithm2e}
	
\usepackage{mathrsfs}

\newtheorem{theorem}{Theorem}

\newtheorem{lemma}{Lemma}

\usepackage{amsmath,amsfonts,amssymb, amsthm, bm}

\usepackage[square,comma,numbers,sort&compress]{natbib}

\title{Carrier phase and amplitude estimation for phase shift keying using pilots and data}
\author{Robby McKilliam, Andr\'{e} Pollok, Bill Cowley, I. Vaughan L. Clarkson and Barry Quinn  
\thanks{
A preliminary version of this paper has been submitted to ICASSP'13~\cite{McKilliam_leastsqPSKnoncoICASSP_2012}.  Supported under the Australian Government’s Australian Space Research Program.
Robby McKilliam, Andr\'{e} Pollok and Bill Cowley are with the Institute for Telecommunications Research, The University of South Australia, SA, 5095.  Vaughan~Clarkson is with the School of Information Technology \& Electrical Engineering, The University of Queensland, QLD., 4072, Australia.  Barry~Quinn is with the Department of Statistics, Macquarie University, Sydney, NSW, 2109, Australia.
}}

\begin{document}

\maketitle

\begin{abstract}
We consider least squares estimators of carrier phase and amplitude from a noisy communications signal that contains both pilot signals, known to the receiver, and data signals, unknown to the receiver.  We focus on signaling constellations that have symbols evenly distributed on the complex unit circle, i.e., $M$-ary phase shift keying.  We show, under reasonably mild conditions on the distribution of the noise, that the least squares estimator of carrier phase is strongly consistent and asymptotically normally distributed.  However, the amplitude estimator is not consistent, but converges to a positive real number that is a function of the true carrier amplitude, the noise distribution and the size of the constellation.  Our theoretical results can also be applied to the case where no pilot symbols exist, i.e., noncoherent detection.  The results of Monte Carlo simulations are provided and these agree with the theoretical results.   
\end{abstract}

\section{Introduction}

In passband communication systems the transmitted signal typically undergoes time offset (delay), phase shift and attenuation (amplitude change).  These effects must be compensated for at the receiver. In this paper we assume that the time offset has been previously handled, and we focus on estimating the phase shift and attenuation.  We consider signalling constellations that have symbols evenly distributed on the complex unit circle such as binary phase shift keying (BPSK), quaternary phase shift keying (QPSK) and $M$-ary phase shift keying ($M$-PSK).  In this case, the transmitted symbols take the form,
\[
s_i = e^{j u_i},
\]
where $j = \sqrt{-1}$ and $u_i$ is from the set $\{0, \tfrac{2\pi}{M}, \dots, \tfrac{2\pi(M-1)}{M}\}$ and $M \geq 2$ is the size of the constellation.  We assume that some of the transmitted symbols are \emph{pilot symbols} known to the receiver and the remainder are information carrying \emph{data symbols} with phase that is unknown to the receiver.  So,
\[
s_i = \begin{cases}
p_i & i \in P \\
d_i & i \in D,
\end{cases}
\]
where $P$ is the set of indices describing the position of the pilot symbols $p_i$, and $D$ is a set of indices describing the position of the data symbols $d_i$.  The sets $P$ and $D$ are disjoint, i.e., $P \cap D = \emptyset$  where $\emptyset$ is the empty set, and we let $L = \abs{P \cup D}$ be the total number of symbols transmitted.

We assume that time offset estimation has been performed and that $L$ noisy $M$-PSK symbols are observed by the receiver.  The received signal after matched filtering is,
\begin{equation}\label{eq:sigmod}
y_i = a_0 s_i + w_i, \qquad i \in P \cup D,
\end{equation}
where $w_i$ is noise and $a_0 = \rho_0 e^{j\theta_0}$ is a complex number representing both carrier phase $\theta_0$ and amplitude $\rho_0$ (by definition $\rho_0$ is a positive real number).  Our aim is to estimate $a_0$ from the noisy symbols $\{ y_i, i \in P \cup D \}$.  Complicating matters is that the data symbols $\{d_i, i \in D\}$ are not known to the receiver and must also be estimated.  Estimation problems of this type have undergone extensive prior study~\cite{ViterbiViterbi_phase_est_1983,Cowley_ref_sym_carr_1998,Wilson1989,Makrakis1990,Liu1991,Mackenthun1994,Sweldens2001,McKilliamLinearTimeBlockPSK2009,Divsalar1990}.  A practical approach is the least squares estimator, that is, the minimisers of the sum of squares function
\begin{equation}\label{eq:SSdefn}
\begin{split}
SS(a, &\{d_i, i \in D\}) = \sum_{i \in P \cup D} \abs{ y_i - a s_i }^2  \\
&= \sum_{i \in P} \abs{ y_i - a p_i }^2 + \sum_{i \in D} \abs{ y_i - a d_i }^2,
\end{split}
\end{equation}
where $\abs{x}$ denotes the magnitude of the complex number $x$.  The least squares estimator is also the maximum likelihood estimator under the assumption that the noise sequence $\{w_i, i \in \ints\}$ is additive white and Gaussian.  However, as we show, the estimator works well under less stringent assumptions.  

The existing literature~\cite{Mackenthun1994,Cowley_ref_sym_carr_1998,ViterbiViterbi_phase_est_1983,Sweldens2001,Wilson1989,Makrakis1990,Liu1991} mostly considers what is called \emph{noncoherent detection} where no pilot symbols exist ($P = \emptyset$).  In the noncoherent setting \emph{differential encoding} is often used, and for this reason the estimation problem has been called \emph{multiple symbol differential detection}.  A popular approach is the so called \emph{non-data aided}, sometimes also called \emph{non-decision directed}, estimator based on the paper of Viterbi and Viterbi~\cite{ViterbiViterbi_phase_est_1983}.  The idea is to `strip' the modulation from the received signal by taking $y_i / \abs{y_i}$ to the power of $M$.  A function $F: \reals \mapsto \reals$ is chosen and the estimator of the carrier phase $\theta_0$ is taken to be $\tfrac{1}{M}\angle{A}$ where $\angle$ denotes the complex argument and
\begin{equation}\label{eq:viterbiviterbi}
A = \frac{1}{L}\sum_{i \in P \cup D} F(\abs{y_i}) \big(\tfrac{y_i}{\abs{y_i}}\big)^M.
\end{equation}
Various choices for $F$ are suggested in~\cite{ViterbiViterbi_phase_est_1983} and a statistical analysis is presented.  A caveat of this estimator is that it is not obvious how pilot symbols should be included. 
This problem does not occur with the least square estimator.

An important paper is by Mackenthun~\cite{Mackenthun1994} who described an algorithm to compute the least squares estimator requiring only $O(L \log L)$ arithmetic operations.  Sweldens~\cite{Sweldens2001} rediscovered Mackenthun's algorithm in 2001.  Both Mackenthun and Swelden considered only the noncoherent setting, but we show in Section~\ref{sec:least-squar-estim}~that Mackenthun's algorithm can be modified to include pilot symbols. Our model includes the noncoherent case by setting the number of pilot symbols to zero, that is, putting $P = \emptyset$.  

In the literature it has been common to assume that the data symbols $\{d_i, i \in D\}$ are of primary interest and that the complex amplitude $a_0$ is a nuisance parameter.  The metric of performance is correspondingly the \emph{symbol error rate}, or \emph{bit error rate}.  While estimating the symbols (or more precisely the transmitted bits) is ultimately the goal, we take the opposite point of view here.  Our aim is to estimate $a_0$, and we treat the unknown data symbols as nuisance parameters.  This is motivated by the fact that in many modern communication systems the data symbols are \emph{coded}.  For this reason raw symbol error rate is not of interest at this stage.  Instead, we desire an accurate estimator $\hat{a}$ of $a_0$, so that the compensated received symbols $\hat{a}^{-1}y_i$ can be accurately modelled using an additive noise channel.  The additive noise channel is a common assumption for subsequent receiver operations, such as decoding.  The estimator $\hat{a}$ is also used in the computation of decoder metrics for modern decoders, and for interference cancellation in multiuser systems.  Consequently, our metric of performance will not be symbol or bit error rate, but $\sabs{\hat{a} - a_0}$. It will be informative to consider the carrier phase and amplitude estimators separately, that is, if $\hat{a} = \hat{\rho}e^{j\hat{\theta}}$ where $\hat{\rho}$ is a positive real number, then we consider $\sabs{\langle\hat{\theta} - \theta_0\rangle_\pi}$ and $\sabs{\hat{\rho} - \rho_0}$.  The function $\fracpart{\cdot}_\pi$ denotes its argument taken `modulo $2\pi$' into the interval $[-\pi, \pi)$.  It will become apparent why $\langle\hat{\theta} - \theta_0\rangle_\pi$ rather than $\hat{\theta} - \theta_0$ is the appropriate measure of error for the phase parameter.

It is possible to generalise the results we present here to allow data symbols with varying constellation size, i.e. varying $M$.  For example, one might give more importance to certain data symbols and use BPSK ($M=2$) for these, but QPSK ($M=4$) for other less important symbols.  This is related to what is called \emph{unequal error protection} in the literature~\cite{Aydinlik_turbo_uep_2008,Sandberg_uep_ldpc_2010}.  To keep our ideas and notation focused we don't consider this further here.

The paper is organised in the following way.  Section~\ref{sec:least-squar-estim} extends Mackenthun's algorithm for the coherent case, when both pilot symbols and data symbols exist.  Section~\ref{sec:circ-symm-compl} describes properties of complex random variables that we need.  Section~\ref{sec:stat-prop-least} states two theorems that describe the statistical properties of the least squares estimator of carrier phase $\hat{\theta}$ and amplitude $\hat{\rho}$.  We show, under some reasonably general assumptions about the distribution of the noise $w_1,\dots,w_L$, that $\langle\hat{\theta} - \theta_0\rangle_\pi$ converges almost surely to zero and that $\sqrt{L}\langle\hat{\theta} - \theta_0\rangle_\pi$ is asymptotically normally distributed as $L\rightarrow \infty$.  However, $\hat{\rho}$ is not a consistent estimator of the amplitude $\rho_0$.  The asymptotic bias of $\hat{\rho}$ is small when the signal to noise ratio (SNR) is large, but the asymptotic bias is significant when the SNR is small.  Sections~\ref{sec:proof-almost-sure} and~\ref{sec:proof-asympt-norm} provide proofs of the theorems stated in Section~\ref{sec:stat-prop-least}.  In Section~\ref{sec:gaussian-noise-case} we consider the special case when the noise is Gaussian.  In this case, our expressions for the asymptotic distribution can be simplified.  Section~\ref{sec:simulations} presents the results of Monte-Carlo simulations.  These simulations agree with the derived asymptotic properties. 


\section{Mackenthun's algorithm with pilots}\label{sec:least-squar-estim}

In this section we derive Mackentun's algorithm to compute the least squares estimator of the carrier phase and amplitude~\cite{Mackenthun1994}.  Mackenthun specifically considered the noncoherent setting, so we modify the algorithm to include the pilot symbols.  For the purpose of analysing computational complexity, we will assume that the number of data symbols $\abs{D}$ is proportional to the total number of symbols $L$, so that, for example, $O(L) = O(\abs{D})$.  In this case Mackentun's algorithm requires $O(L \log L)$ arithmetic operations.  This complexity arises from the need to sort a list of $\abs{D}$ elements.  


Define the sum of squares function
\begin{align}
SS(a, &\{d_i, i \in D\}) = \sum_{i \in P \cup D} \abs{ y_i - a s_i }^2 \nonumber \\
&= \sum_{i \in P \cup D} \big( \abs{y_i}^2 - a s_i y_i^* - a^* s_i^* y_i + aa^* \big), \label{eq:SS}
\end{align}
where $*$ denotes complex conjugate.  The minimiser of $SS$ with respect to $a$ as a function of $\{d_i, i \in D\}$ is
\begin{equation}\label{eq:hata}
\hat{a}(\{d_i, i \in D\}) = \frac{1}{L} \sum_{i \in P \cup D} y_i s_i^* = \frac{1}{L} Y
\end{equation}
where $L = \abs{P \cup D}$ is the total number of symbols transmitted, and to simplify our notation we have put 
\[
Y = \sum_{i \in P \cup D} y_i s_i^* = \sum_{i \in P } y_i p_i^* + \sum_{i \in D } y_i d_i^*.
\]  
Note that $Y$ is a function of the unknown data symbols $\{ d_i, i \in D\}$ and we could write $Y(\{ d_i, i \in D\})$, but have chosen to suppress the argument $(\{ d_i, i \in D\})$ for notational brevity.  Substituting $\frac{1}{L}Y$ for $a$ into~\eqref{eq:SS} we obtain $SS$ minimised with respect to $a$,
\begin{equation}\label{eq:SSdatasymbols}
SS(\{d_i, i \in D\}) = A - \frac{1}{L}\abs{Y}^2,
\end{equation}
where $A = \sum_{i \in P \cup D}\abs{y_i}^2$ does not depend on the $d_i$.  The least squares estimators of the data symbols are the minimisers of~\eqref{eq:SSdatasymbols}.  Observe that given candidate values for the data symbols, we can compute the corresponding $SS(\{d_i, i \in D\})$ in $O(L)$ arithmetic operations.  It turns out that there are at most $M\abs{D}$ candidate values of the least squares estimator of the data symbols~\cite{Sweldens2001,Mackenthun1994}.  

To see this, let $a = \rho e^{j\theta}$ where $\rho$ is a nonnegative real.  Now,
\begin{align}
SS(\rho, \theta, &\{d_i, i \in D\}) = \sum_{i \in P \cup D} \abs{ y_i - \rho e^{j\theta} s_i }^2  \nonumber \\
&= \sum_{i \in P} \abs{ y_i - \rho e^{j\theta} p_i }^2 + \sum_{i \in D} \abs{ y_i - \rho e^{j\theta} d_i }^2. \label{eq:SSallparams}
\end{align}
We have slightly abused notation here by reusing $SS$. This should not cause confusion as $SS(a, \{d_i, i \in D\})$, $SS(\rho, \theta, \{d_i, i \in D\})$, and $SS(\{d_i, i \in D\})$ are easily told apart by their arguments.  For given $\theta$, the least squares estimator of the $i$th data symbol $d_i$ is given by minimising $\abs{ y_i - \rho e^{j\theta} d_i }^2$, that is,
\begin{equation}\label{eq:hatdfinxtheta}
\hat{d}_i(\theta) = e^{j\hat{u}_i(\theta)} \qquad \text{where} \qquad \hat{u}_i(\theta) = \round{\angle( e^{-j\theta}y_i)},
\end{equation}
where $\angle(\cdot)$ denotes the complex argument (or phase), and $\round{\cdot}$ rounds its argument to the nearest multiple of $\frac{2\pi}{M}$.  A word of caution, the notation $\round{\cdot}$ is often used to denote rounding to the nearest \emph{integer}.  This is not the case here.  If the function $\operatorname{round}(\cdot)$ takes its argument to the nearest integer then,
\[
\round{x} = \tfrac{2\pi}{M}\operatorname{round}\left(\tfrac{M}{2\pi}x\right).
\] 
Note that $\hat{d}_i(\theta)$ does not depend on $\rho$.  As defined, $\hat{u}_i(\theta)$ is not strictly inside the set $\{0, \tfrac{2\pi}{M}, \dots, \tfrac{2\pi(M-1)}{M}\}$, but this is not of consequence, as we intend its value to be considered equivalent modulo $2\pi$.  With this in mind,
\[
\hat{u}_i(\theta) = \round{\angle{y_i} - \theta }
\]
which is equivalent to the definition from~\eqref{eq:hatdfinxtheta} modulo $2\pi$.

We only require to consider $\theta$ in the interval $[0, 2\pi)$.  Consider how $\hat{d}_i(\theta)$ changes as $\theta$ varies from $0$ to $2\pi$.  Let $b_i = \hat{d}_i(0)$ and let 
\[
z_i = \angle{y_i} - \hat{u}_i(0) = \angle{y_i} - \round{\angle{y_i}}
\]
be the phase difference between the received symbol $y_i$ and the hard decision resulting when $\theta = 0$, i.e. $\round{\angle{y_i}}$.  Then,
\begin{equation}\label{eq:uicombos}
\hat{d}_i(\theta) = 
\begin{cases}
b_i, &  0 \leq \theta < z_i + \frac{\pi}{M} \\
b_i e^{-j2\pi/M}, & z_i + \frac{\pi}{M} \leq \theta < z_i + \frac{3\pi}{M} \\ 
\vdots & \\
b_i e^{-j2\pi k /M}, & z_i + \frac{\pi(2k - 1)}{M} \leq \theta < z_i + \frac{\pi(2k + 1)}{M}  \\ 
\vdots & \\
b_i e^{-j2\pi} = b_i, &  z_i + \frac{\pi(2M - 1)}{M} \leq \theta < 2\pi. \\
\end{cases}
\end{equation}

Let 
\[
f(\theta) = \{ \hat{d}_i(\theta), i \in D \}
\]
be a function mapping the interval $[0, 2\pi)$ to a sequence of $M$-PSK symbols indexed by the elements of $D$.  Observe that $f(\theta)$ is piecewise continuous.  The subintervals of $[0, 2\pi)$ over which $f(\theta)$ remains constant are determined by the values of $\{z_i, i \in D\}$.  Let
\[
S = \{ f(\theta) \mid \theta \in [0, 2 \pi) \}
\]
be the set of all sequences $f(\theta)$ as $\theta$ varies from $0$ to $2\pi$.  If $\hat{\theta}$ is the least squares estimator of the phase then $S$ contains the sequence $\{ \hat{d}_i(\hat{\theta}), i \in D \}$ corresponding to the least squares estimator of the data symbols, i.e., $S$ contains the minimiser of~\eqref{eq:SSdatasymbols}.  Observe from~\eqref{eq:uicombos} that there are at most $M\abs{D}$ sequences in $S$, because there are $M$ distinct values of $d_i(\theta)$ for each $i \in D$ as $\theta$ varies from $0$ to $2\pi$.

The sequences in $S$ can be enumerated as follows.  Let $\sigma$ denote the permutation of the indices in $D$ such that $z_{\sigma(i)}$ are in ascending order, that is,
\begin{equation}\label{eq:sigmasortind}
z_{\sigma(i)} \leq z_{\sigma(k)}
\end{equation}
whenever $i < k $ where $i, k \in \{0, 1, \dots, \abs{D}-1\}$.  It is convenient to define the indices into $\sigma$ to be taken modulo $\abs{D}$, that is, if $m$ is an integer not from $\{0, 1, \dots, \abs{D}-1\}$ then we define $\sigma(m) = \sigma(k)$ where $k \equiv m \mod \abs{D}$ and $k \in  \{0, 1, \dots, \abs{D}-1\}$.  The first sequence in $S$ is 
\[
f_0 = f(0) = \{ \hat{d}_i(0), i \in D \} = \{ b_i, i \in D \}.
\]  
The next sequence $f_1$ is given by replacing the element $b_{\sigma(0)}$ in $f_0$ with $b_{\sigma(0)}e^{-j2\pi/M}$.  Given a sequence $x$ we use $x e_i$ to denote $x$ with the $i$th element replaced by $x_i e^{-j2\pi/M}$.  Using this notation,  
\[
f_1 = f_0 e_{\sigma(0)}.
\] 
The next sequence in $S$ is correspondingly 
\[
f_2 = f_0 e_{\sigma(0)} e_{\sigma(1)} = f_1 e_{\sigma(1)},
\]
and the $k$th sequence is
\begin{equation}\label{eq:fkrec}
f_{k+1} = f_{k} e_{\sigma(k)}.
\end{equation}
In this way, all $M\abs{D}$ sequences in $S$ can be recursively enumerated.

We want to find the $f_k \in S$ corresponding to the minimiser of~(\ref{eq:SSdatasymbols}).  A na\"{\i}ve approach would be to compute $SS(f_k)$ for each $k \in \{0,1,\dots,M\abs{D}-1\}$.  Computing $SS(f_k)$ for any particular $k$ requires $O(L)$ arithmetic operations.  So, this na\"{\i}ve approach would require $O(L M \abs{D}) = O(L^2)$ operations in total.  Following Mackenthun~\cite{Mackenthun1994}, we show how $SS(f_k)$ can be computed recursively.

Let,
\begin{equation}\label{eq:SSfk}
SS(f_k) = A - \frac{1}{L}\abs{Y_k}^2,
\end{equation}
where, 
\begin{align*}
Y_k = Y( f_k ) &= \sum_{i \in P} y_i p_i^*  + \sum_{i \in D} y_i f_{ki}^* \\
&= B + \sum_{i \in D}g_{ki},
\end{align*}
where $B = \sum_{i \in P} y_i p_i^*$ is independent of the data symbols, and $f_{ki}$ denotes the $i$th symbol in $f_k$, and for convenience, we put $g_{ki}  = y_i f_{ki}^*$.  Letting $g_{k}$ be the sequence $\{g_{ik}, i \in D\}$ we have, from~\eqref{eq:fkrec}, that $g_k$ satisfies the recursive equation
\[
g_{k+1} = g_{k} e_{\sigma(k)}^*,
\]
where $g_{k} e_{\sigma(k)}^*$ indicates the sequence $g_k$ with the $\sigma(k)$th element replaced by $g_{k \sigma(k)}e^{j2\pi/M}$.  Now,
\[
Y_0 = B + \sum_{i \in D} g_{0i}
\] 
can be computed in $O(L)$ operations, and
\begin{align*}
Y_1 &= B + \sum_{i \in D} g_{1i} \\
&= B +  (e^{j2\pi/M} - 1)g_{0\sigma(0)} + \sum_{i \in D} g_{0i} \\
&= Y_0 + \eta g_{0\sigma(0)},
\end{align*}
where $\eta = e^{j2\pi/M} - 1$.  In general,
\[
Y_{k+1} = Y_k + \eta g_{k\sigma(k)}.
\]
So, each $Y_k$ can be computed from it predecessor $Y_{k-1}$ in a $O(1)$ arithmetic operations.  Given $Y_k$, the value of $SS(f_k)$ can be computed in $O(1)$ operations using~\eqref{eq:SSfk}.  Let $\hat{k} = \arg\min SS(f_k)$.  The least squares estimator of $a_0$ is then computed according to~\eqref{eq:hata},
\begin{equation}\label{eq:ahatYhat}
\hat{a} = \frac{1}{L} Y_{\hat{k}}.
\end{equation}
Pseudocode is given in Algorithm~\ref{alg:loglinear}.  Line~\ref{alg_sortindices} contains the function $\operatorname{sortindices}$ that, given $z = \{z_i, i \in D\}$, returns the permutation $\sigma$ as described in~\eqref{eq:sigmasortind}.  The $\operatorname{sortindicies}$ function requires sorting $\abs{D}$ elements.  This requires $O(L \log L)$ operations.  The $\operatorname{sortindicies}$ function is the primary bottleneck in this algorithm when $L$ is large.  The loops on lines~\ref{alg_loop_setup} and~\ref{alg_loop_search} and the operations on lines~\ref{alg_Y} to lines~\ref{alg_Q} all require $O(L)$ or less operations.  

\begin{algorithm}[t] \label{alg:loglinear}
\SetAlCapFnt{\small}
\SetAlTitleFnt{}
\caption{Mackenthun's algorithm with pilot symbols}
\DontPrintSemicolon
\KwIn{$\{y_i, i \in P \cup D \}$}
\For{$i \in D$ \nllabel{alg_loop_setup}}{
$\phi = \angle{y_i}$ \;
$u = \round{\phi} $ \;
$z_i = \phi -  u $ \;
$g_i = y_i e^{-j u}$ \nllabel{alg_gset} \;
}
$Y = \sum_{i \in P} y_i p_i^* + \sum_{i \in D} g_i $ \nllabel{alg_Y}\;
$\hat{a} = \frac{1}{L} Y$ \nllabel{alg_hata1} \;
$\hat{Q} = \frac{1}{L}\abs{Y}^2$ \nllabel{alg_Q} \;
$\eta = e^{j2\pi/M} - 1$ \;
$\sigma = \operatorname{sortindices}(z)$ \nllabel{alg_sortindices} \;
\For{$k= 0$ \emph{\textbf{to}} $M\abs{D}-1$ \nllabel{alg_loop_search}}{
$Y = Y + \eta g_{\sigma(k)}$ \;
$g_{\sigma(k)} = (\eta + 1) g_{\sigma(k)} $\;
$Q = \frac{1}{L}\abs{Y}^2$\;
\If{$Q > \hat{Q}$}{
 	$\hat{Q} = Q$ \;
 	$\hat{a} =  \frac{1}{L} Y$ \nllabel{alg_hata2} \;
 }
}
\Return{$\hat{a}$ \nllabel{alg_return}}
\end{algorithm}

\section{Circularly symmetric complex random variables}\label{sec:circ-symm-compl}

Before describing the statistical properties of the least squares estimator, we first require some properties of complex valued random variables.  A complex random variable $W$ is said to be \emph{circularly symmetric} if its phase $\angle{W}$ is independent of its magnitude $\abs{W}$ and if the distribution of $\angle{W}$ is uniform on $[0,2\pi)$.  That is, if $Z \geq 0$ and $\Theta \in [0,2\pi)$ are real random variables such that $Ze^{j\Theta} = W$, then $\Theta$ is uniformly distributed on $[0,2\pi)$ and is independent of $Z$.  If the probability density function (pdf) of $Z$ is $f_Z(z)$, then the joint pdf of $\Theta$ and $Z$ is 
\[
f_{Z,\Theta}(z,\theta) = \frac{1}{2\pi}f_Z(z).
\]
Observe that for any real number $\phi$, the pdf of $W$ and $e^{j\phi}W$ are the same, that is, the pdf is invariant to phase rotation.  If $\expect\abs{W} = \expect Z$ is finite, then $W$ has zero mean because
\begin{align*}
 \expect W &= \int_{0}^{2\pi} \int_{0}^\infty z e^{j\theta} f_{Z,\Theta}(z,\theta) dz d\theta \\
 &= \frac{1}{2\pi} \int_{0}^{2\pi} e^{j\theta} \int_{0}^\infty z f_Z(z) dz d\theta \\
 &= \frac{1}{2\pi}\expect Z \int_{0}^{2\pi} e^{j\theta} d\theta = 0.
 \end{align*}
If $X$ and $Y$ are real random variables equal to the real and imaginary parts of $W = X + jY$ then the joint pdf of $X$ and $Y$ is
\[
f_{X,Y}(x,y) = \frac{f_Z(\sqrt{x^2 + y^2})}{2\pi \sqrt{x^2 + y^2}}.
\]

We will have particular use of complex random variables of the form $1 + W$ where $W$ is circularly symmetric.  Let $R \geq 0$ and $\Phi \in [0,2\pi)$ be real random variables satisfying, 
\[
R e^{j\Phi} = 1 + W.
\]
The joint pdf of $R$ and $\Phi$ can be shown to be
\begin{equation}\label{eq:pdfRPhi}
f(r,\phi) = \frac{r f_Z(\sqrt{r^2 - 2r\cos\phi + 1})}{2\pi\sqrt{r^2 - 2r\cos\phi + 1}}.
\end{equation}
Since $\cos\phi$ has period $2\pi$ and is even on $[-\pi,\pi]$ it follows that $f(r,\phi)$ has period $2\pi$ and is even on $[-\pi,\pi]$ with respect to $\phi$.  The mean of $R e^{j\Phi}$ is equal to one because the mean of $W$ is zero.  So,
\begin{equation}\label{eq:expectRepartRphi}
\expect \Re(R e^{j\Phi}) = \expect R \cos(\Phi) = 1,
\end{equation}
where $\Re(\cdot)$ denotes the real part, and
\begin{equation}\label{eq:expectImpartRphi}
\expect \Im(R e^{j\Phi}) = \expect R \sin(\Phi) = 0,
\end{equation}
where $\Im(\cdot)$ denotes the imaginary part.



\section{Statistical properties of the least squares estimator}\label{sec:stat-prop-least}

In this section we describe the asymptotic properties of the least squares estimator.  In what follows we use $\sfracpart{x}_\pi$ to denote $x$ taken `modulo $2\pi$' into the interval $[-\pi, \pi)$, that is
\[
\fracpart{x}_\pi = x - 2\pi\operatorname{round}\left(\frac{x}{2\pi}\right),
\]
where $\operatorname{round}(\cdot)$ takes its argument to the nearest integer.  The direction of rounding for half-integers is not important so long as it is consistent.  We have chosen to round up half-integers here.  Similarly we use $\sfracpart{x}$ to denote $x$ taken `modulo $\tfrac{2\pi}{M}$' into the interval $\left[-\tfrac{\pi}{M}, \tfrac{\pi}{M}\right)$, that is
\[
\fracpart{x} = x - \tfrac{2\pi}{M}\operatorname{round}\left(\tfrac{M}{2\pi}x\right) = x - \round{x}.
\]
The next two theorems describe the asymptotic properties of the least squares estimator.  These are the central results and the chief original contributions of this paper.

\begin{theorem}\label{thm:consistency} (Almost sure convergence)
Let $\{w_i\}$ be a sequence of independent and identically distributed, circularly symmetric complex random variables with $\expect \abs{w_1}^2$ finite, and let $\{y_i, i \in P \cup D\}$ be given by~\eqref{eq:sigmod}.   Let $\hat{a} = \hat{\rho}e^{j\hat{\theta}}$ be the least squares estimator of $a_0 = \rho_0e^{j\theta_0}$. 
Put $L = \abs{P \cup D}$ and let $\abs{P}$ and $\abs{D}$ increase in such a way that
\[
\frac{\abs{P}}{L} \rightarrow p \qquad \text{and} \qquad \frac{\abs{D}}{L} \rightarrow d \qquad \text{as $L \rightarrow \infty$.}
\] 
Let $R_i \geq 0$ and $\Phi_i \in [0,2\pi)$ be real random variables satisfying
\begin{equation}\label{eq:RiandPhii}
R_ie^{j\Phi_i} = 1 + \frac{w_i}{a_0 s_i} ,
\end{equation}
and define the continuous function
\[
G(x) = p h_1(x) + d h_2(x) \qquad \text{where}
\]
\[
h_1(x) = \expect R_1 \cos(x + \Phi_1), \;\;\; h_2(x) =  \expect R_1 \cos\sfracpart{ x + \Phi_1}.
\]
If $p > 0$ and if $G(x)$ is uniquely maximised at $x = 0$ over the interval $[-\pi,\pi)$ then
\begin{enumerate}
\item $\sfracpart{\hat{\theta} - \theta_0}_\pi \rightarrow 0$ almost surely as $L \rightarrow \infty$,
\item $\hat{\rho} \rightarrow \rho_0 G(0)$ almost surely as $L \rightarrow \infty$.
\end{enumerate}
\end{theorem}

\begin{theorem}\label{thm:normality} (Asymptotic normality)
Under the same conditions as Theorem~\ref{thm:consistency}, let $f(r,\phi)$ be the joint probability density function of $R_1$ and $\Phi_1$, let
\[
g(\phi) = \int_{0}^{\infty} r f(r,\phi) dr 
\]
and assume that $\frac{\abs{P}}{L} = p + o(L^{-1/2})$ and $\frac{\abs{D}}{L} = d + o(L^{-1/2})$.
Put $\hat{\lambda}_L = -\sfracpart{\hat{\theta} - \theta_0}_\pi = \sfracpart{\theta_0 - \hat{\theta}}_\pi$ and $\hat{m}_L = \hat{\rho} - \rho_0 G(0)$. 
If the function $g$ is continuous at $\tfrac{2\pi}{M}k + \tfrac{\pi}{M}$ for each $k = 0, \dots, M-1$, then the distribution of $(\sqrt{L}\hat{\lambda}_L, \sqrt{L}\hat{m}_L)$ converges to the bivariate normal with zero mean and covariance matrix
\[
\left( \begin{array}{cc} 
\frac{pA_1 + dA_2}{(p + H d)^2} & 0 \\
0 & \rho_0^2(pB_1 + dB_2)
\end{array} \right)
\]
as $L \rightarrow \infty$, where
\[
H = h_2(0) -  2 \sin(\tfrac{\pi}{M}) \sum_{k = 0}^{M-1} g(\tfrac{2\pi}{M}k + \tfrac{\pi}{M}),
\]
\[
A_1 = \expect R_1^2\sin^2(\Phi_1), \qquad A_2 = \expect R_1^2\sin^2\fracpart{\Phi_1},
\]
\[
B_1 = \expect R_1^2 \cos^2(\Phi_1) - 1, \;\;\; B_2 = \expect R_1^2 \cos^2\fracpart{\Phi_1} - h_2^2(0).
\]
\end{theorem}


The proof of Theorem~\ref{thm:consistency} is in Section~\ref{sec:proof-almost-sure} and the proof of Theorem~\ref{thm:normality} is in Section~\ref{sec:proof-asympt-norm}.  Before giving the proofs we discuss the assumptions made by these theorems.  The assumption that $w_1, \dots, w_L$ are circularly symmetric can be relaxed, but this comes at the expense of making the theorem statements more complicated.  If $w_i$ is not circularly symmetric then the distribution of $R_i$ and $\Phi_i$ may depend on $a_0$ and also on the transmitted symbols $\{s_i,i \in P \cup D\}$.  As a result the asymptotic variance described in Theorem~\ref{thm:normality} depends on $a_0$ and $\{s_i,i \in P \cup D\}$, rather than just $\rho_0$.  The circularly symmetric assumption may not always hold in practice, but we feel it provides a sensible trade off between simplicity and generality.  

The assumption that $\expect \sabs{w_1}^2 = \expect \sabs{w_i}^2$ is finite implies that $R_i$ has finite variance since $\expect R_i^2 = 1 + \expect \abs{w_i}^2$.  This is required in Theorem~\ref{thm:normality} so that the constants $A_1$, $A_2$, $B_1$ and $B_2$ exist.  We will also use the fact that $R_i$ has finite variance to simplify the proof of Theorem~\ref{thm:consistency} by use of Kolmogorov's strong law of large numbers~\cite{Billingsley1979_probability_and_measure}.

The theorems place conditions on $\sfracpart{\hat{\theta} - \theta_0}_\pi$ rather than directly on $\hat{\theta}  - \theta_0$.  This makes sense because the phases $\theta_0$ and $\theta_0 + 2\pi k$ are equivalent for any integer $k$. So, for example, we expect the phases $0.99\pi$ and $-0.99\pi$ to be close together, the difference between them being $\vert\sfracpart{-0.99\pi - 0.99\pi}_\pi\vert = 0.02\pi$, and not $\vert -0.99\pi - 0.99\pi\vert = 1.98\pi$.

Theorem~\ref{thm:normality} requires the function $g$ to be continuous at $\tfrac{2\pi}{M}k + \tfrac{\pi}{M}$ for each $k = 0, \dots, M-1$.  This places mild restrictions on the distribution of the noise $w_i$.  For example, the requirements are satisfied if the joint pdf of the real and imaginary parts of $w_i$ is continuous, since in this case $f(r,\phi)$ is continuous.  Because $f(r,\phi)$ has period $2\pi$ and is even on $[-\pi,\pi]$ with respect to $\phi$ it follows that $g$ has period $2\pi$ and is even on $[-\pi, \pi]$.



A key assumption in Theorem~\ref{thm:consistency} is that $G(x)$ is uniquely maximised at $x = 0$ for $x \in [-\pi, \pi)$.  This assumption asserts that $G(x) \leq G(0)$ for all $x \in [-\pi, \pi)$ and that if $\{x_i\}$ is a sequence of numbers from $[-\pi,\pi)$ such that $G(x_i) \rightarrow G(0)$ as $i \rightarrow \infty$ then $x_i \rightarrow 0$ as $i \rightarrow \infty$.  Although we will not prove it here, this assumption is not only sufficient, but also necessary, for if $G(x)$ is uniquely maximised at some $x \neq 0$ then $\sfracpart{\hat{\theta} - \theta_0}_\pi \rightarrow x$ almost surely as $L\rightarrow\infty$, while if $G(x)$ is not uniquely maximised then $\sfracpart{\hat{\theta} - \theta_0}_\pi$ will not converge.  One can check that this assumption holds when $w_1$ is circularly symmetric and normally distributed.  We will not attempt to further classify those distributions for which the assumption holds here.



Theorem~\ref{thm:consistency} defines real numbers $p$ and $d$ to represent the proportion of pilot symbols and data symbols in the limit as $L$ goes to infinity.  For Theorem~\ref{thm:normality} we need the slightly stronger condition that 
\[
\frac{\abs{P}}{L} = p + o(L^{-1/2}) \qquad \text{and} \qquad \frac{\abs{D}}{L} = d + o(L^{-1/2}).
\] 
This stronger condition is required to prove the asymptotic normality of $\sqrt{L}\hat{m}_L$.   

The next two sections give proofs of Theorems~\ref{thm:consistency}~and~\ref{thm:normality}.  The proofs make use of various lemmas, which are proved in the appendix.


\section{Proof of almost sure convergence (Theorem~\ref{thm:consistency}) } \label{sec:proof-almost-sure}

Substituting $\{ \hat{d}_i(\theta), i \in D \}$ from~\eqref{eq:hatdfinxtheta} into~\eqref{eq:SSallparams} we obtain $SS$ minimised with respect to the data symbols,
 \begin{align*}
SS(\rho, \theta) &=\sum_{i \in P} \abs{ y_i - \rho e^{j\theta} p_i }^2 + \sum_{i \in D} \abs{ y_i - \rho e^{j\theta} \hat{d_i}(\theta) }^2 \\
&= A - \rho Z(\theta) - \rho Z^*(\theta) + L \rho^2,
\end{align*}
where
\[
Z(\theta)  = \sum_{i \in P} y_i e^{-j\theta} p_i^* + \sum_{i \in D} y_i e^{-j\theta} \hat{d}_i^*(\theta),
\]
and $Z^*(\theta)$ is the conjugate of $Z(\theta)$.  Differentiating with respect to $\rho$ and setting the resulting expression to zero gives the least squares estimator of $\rho_0$ as a function of $\theta$,
\begin{equation}\label{eq:hatrhoZ}
\hat{\rho}(\theta) = \frac{Z(\theta) + Z^*(\theta)}{2L} = \frac{1}{L}\Re(Z(\theta)),
\end{equation}
where $\Re(\cdot)$ denotes the real part.  
Substituting this expression into $SS(\rho, \theta)$ gives $SS$ minimised with respect to $\rho$ and the data symbols,
\[
SS(\theta) = A - \frac{1}{L}\Re(Z(\theta))^2.
\]
We again abuse notation by reusing $SS$, but this should not cause confusion as $SS(\rho,\theta)$ and $SS(\theta)$ are easily told apart by their inputs.  By definition the amplitude $\rho_0$ and its estimator $\hat{\rho}$ are positive.  However, $\hat{\rho}(\theta) = \Re(Z(\theta))$ may take negative values for some $\theta \in [-\pi,\pi)$.  
The least square estimator $\hat{\theta}$ of $\theta_0$ is the minimiser of $SS(\theta)$ under the constraint $\hat{\rho}(\theta) = \Re(Z(\theta)) > 0$.  Equivalently $\hat{\theta}$ is the maximiser of $\Re(Z(\theta))$ with no constraints required.
 
We are thus interested in analysing the behaviour of the maximiser of $\Re(Z(\theta))$.  Recalling the definition of $R_i$ and $\Phi_i$ from~\eqref{eq:RiandPhii},
\begin{align*}
y_i &= a_0 s_i + w_i \\
&= a_0 s_i \left( 1 + \frac{w_i}{a_0 s_i} \right) \\
&= a_0 s_i R_i e^{j \Phi_i} \\
&= \rho_0 R_i e^{j ( \Phi_i + \theta_0 + \angle{s_i}) }.
\end{align*}
Recalling the definition of $\hat{d}_i(\theta)$ and $\hat{u}_i(\theta)$ from~\eqref{eq:hatdfinxtheta},
\begin{align*}
\hat{u}_i(\theta) &= \round{\angle{y_i} - \theta} \\
&= \round{\theta_0 + \Phi_i + \angle{s_i} - \theta} \\
&\equiv \round{ \fracpart{\theta_0 - \theta}_{\pi} + \Phi_i + \angle{s_i}} \pmod{2\pi} \\
&= \round{ \lambda + \Phi_i + \angle{s_i} },
\end{align*}
where we put $\lambda = \fracpart{\theta_0 - \theta}_\pi$ and where, as in Section~\ref{sec:least-squar-estim}, we consider $\hat{u}_i(\theta)$ equivalent modulo $2\pi$.  Because $\hat{d}_i^*(\theta) = e^{-j\hat{u}_i(\theta)}$, it follows that, when $i \in D$,
\begin{align}
 y_i e^{-j\theta} \hat{d}_i^*(\theta) &= \rho_0 R_i e^{j(\lambda + \Phi_i + \angle{s_i} - \round{\lambda + \Phi_i + \angle{s_i}})} \nonumber \\
&= \rho_0 R_i e^{j(\lambda + \Phi_i - \round{\lambda + \Phi_i})} \nonumber  \\
&= \rho_0 R_i e^{j\sfracpart{\lambda + \Phi_i}} \label{eq:yethetadhat}
\end{align}
since $\round{x + \angle{s_i}} = \round{x} + \angle{s_i}$ for all $x \in \reals$ as a result of $\angle{s_i}$ being a multiple of $\tfrac{2\pi}{M}$.  Otherwise, when $i \in P$,  
\[
y_i e^{-j\theta} p_i^* = \rho_0 R_i e^{j(\lambda + \Phi_i)}.
\]
Now,
\[
Z(\theta) = \rho_0 \sum_{i \in P} R_i e^{j(\lambda + \Phi_i)} + \rho_0  \sum_{i \in D} R_i e^{j\sfracpart{\lambda + \Phi_i}}.
\]
Let 
\begin{equation}\label{eq:GLdefn}
G_L(\lambda) = \frac{1}{\rho_0 L} \Re(Z(\theta))
\end{equation}
and put $\hat{\lambda}_L = -\sfracpart{\hat{\theta} - \theta_0}_\pi = \sfracpart{\theta_0 - \hat{\theta}}_\pi$.  Since $\hat{\theta}$ is the maximiser of $\Re(Z(\theta))$ it follows that $\hat{\lambda}_L$ is the maximiser of $G_L(\lambda)$.  We will show that $\hat{\lambda}_L$ converges almost surely to zero as $L \rightarrow \infty$.  The proof of part 1 of Theorem~\ref{thm:consistency} follows from this.

Recall the functions $G$, $h_1$ and $h_2$ defined in the statement of Theorem~\ref{thm:consistency}.  Observe that
\[
\expect G_L(\lambda) = \frac{\abs{P}}{L} h_1(\lambda) + \frac{\abs{D}}{L} h_2(\lambda)
\]
and since $\frac{\abs{P}}{L} \rightarrow p$ and $\frac{\abs{D}}{L} \rightarrow d$ as $L \rightarrow \infty$,
\[
\lim_{L \rightarrow\infty} \expect G_L(\lambda) = G(\lambda) = p h_1(\lambda)   +  d h_2(\lambda).
\]
As is customary, let $\Omega$ be the sample space on which the random variables $\{w_i\}$ are defined.  Let $A$ be the subset of the sample space $\Omega$ on which $G(\hat{\lambda}_L) \rightarrow G(0)$ as $L\rightarrow\infty$.  Lemma~\ref{lem:convtoexpGlamL} shows that $\prob\{A\} = 1$.  Let $A'$ be the subset of the sample space on which $\hat{\lambda}_L \rightarrow 0$ as $L\rightarrow \infty$.  Because $G(x)$ is uniquely maximised at $x=0$, it follows that $G(\hat{\lambda}_L) \rightarrow G(0)$ only if $\hat{\lambda}_L \rightarrow 0$ as $L \rightarrow\infty$. So $A \subseteq A'$ and therefore $\prob\{A'\} \geq \prob\{A\} = 1$.  Part 1 of Theorem~\ref{thm:consistency} follows.  

It remains to prove part 2 of the theorem regarding the convergence of the amplitude estimator $\hat{\rho}$.  From~\eqref{eq:hatrhoZ},
\begin{equation}\label{eq:rhoGLZ}
\hat{\rho} = \frac{1}{L}\Re(Z(\hat{\theta})) = \rho_0 G_L(\hat{\lambda}_L).
\end{equation}  
Lemma~\ref{lem:GLtoG0} in the appendix shows that $G_L(\hat{\lambda}_L)$ converges almost surely to $G(0)$ as $L\rightarrow\infty$, and $\hat{\rho}$ consequently converges almost surely to $\rho_0 G(0)$ as required.  It remains to prove Lemmas~\ref{lem:convtoexpGlamL} and~\ref{lem:GLtoG0}.  These are proved in Section~\ref{sec:proof-almost-sureappendix} of the appendix.

\section{Proof of asymptotic normality (Theorem~\ref{thm:normality}) } \label{sec:proof-asympt-norm}

We first prove the asymptotic normality of $\sqrt{L} \hat{\lambda}_L$.  Once this is done we will be able to prove the normality of $\sqrt{L} \hat{m}_L$.  Recall that $\hat{\lambda}_L$ is the maximiser of the function $G_L$ defined in~\eqref{eq:GLdefn}.  The proof is complicated by the fact that $G_L$ is not differentiable everywhere due to the function $\fracpart{\cdot}$ not being differentiable at multiples of $\tfrac{\pi}{M}$.  This prevents the use of ``standard approaches'' to proving normality that are based on the mean value theorem~\cite{vonMises_diff_stats_1947,vanDerVart1971_asymptotic_stats,Pollard_new_ways_clts_1986,Pollard_conv_stat_proc_1984,Pollard_asymp_empi_proc_1989,van2009empirical}.  However, Lemma~\ref{lem:diffatlambdaL} shows that the derivative $G_L^\prime$ does exist, and is equal to zero, at $\hat{\lambda}_L$.  Similar properties have been used by some of the present authors to analyse the behaviour of polynomial-phase estimators~\cite{McKilliam_LSU_polyest_part_arxiv_2012}.  
Define the function
\begin{equation}\label{eq:RLdef}
R_L(\lambda) = \frac{1}{L} \sum_{i \in P} R_i \sin(\lambda + \Phi_i) + \frac{1}{L} \sum_{i \in D} R_i \sin\sfracpart{\lambda + \Phi_i}.
\end{equation}
Whenever $G_L(\lambda)$ is differentiable $G_L^\prime(\lambda) = R_L(\lambda)$, and so $R_L(\hat{\lambda}_L) = G_L^\prime(\hat{\lambda}_L) = 0$ by Lemma~\ref{lem:diffatlambdaL}.  Let $Q_L(\lambda) = \expect R_L(\lambda) - \expect R_L(0)$ and write
\begin{align*}
0 &= R_L(\hat{\lambda}_L) - Q_L(\hat{\lambda}_L) + Q_L(\hat{\lambda}_L) \\
&= \sqrt{L}\big( R_L(\hat{\lambda}_L) - Q_L(\hat{\lambda}_L) \big) + \sqrt{L}Q_L(\hat{\lambda}_L).
\end{align*}
Lemma~\ref{lem:Qconv} shows that
\[
\sqrt{L} Q_L(\hat{\lambda}_L) = \sqrt{L} \hat{\lambda}_L\big( p + Hd  + o_P(1) \big)
\]
where $o_P(1)$ denotes a sequence of random variables converging in probability to zero as $L \rightarrow \infty$, and $p,d$ and $H$ are defined in the statement of Theorems~\ref{thm:consistency}~and~\ref{thm:normality}.  Lemma~\ref{lem:empiricprocc} shows that
\[
\sqrt{L}\big( R_L(\hat{\lambda}_L) - Q_L(\hat{\lambda}_L) \big) = o_P(1) + \sqrt{L} R_L(0).
\]
It follows from the three equations above that,
\[
0 = o_P(1) + \sqrt{L}R_L(0) + \sqrt{L} \hat{\lambda}_L \big( p + Hd  + o_P(1) \big)
\]
and rearranging gives,
\[
\sqrt{L} \hat{\lambda}_L = o_P(1) - \frac{\sqrt{L}R_L(0)}{p + Hd  + o_P(1)}.
\]
Lemma~\ref{lem:convdistGLdash} shows that the distribution of $\sqrt{L}R_L(0)$ converges to the normal with zero mean and variance $pA_1 + dA_2$ where $A_1$ and $A_2$ are defined in the statement of Theorem~\ref{thm:normality}.  It follows that the distribution of $\sqrt{L}\hat{\lambda}_L$ converges to the normal with zero mean and variance
\[
\frac{pA_1 + dA_2}{(p + Hd)^2}.
\]
 
We now analyse the asymptotic distribution of $\sqrt{L} \hat{m}_L$.  Let $T_L(\lambda) = \expect G_L(\lambda)$.  Using~\eqref{eq:rhoGLZ},
\begin{align*}
\sqrt{L} \hat{m}_L &= \sqrt{L} \rho_0 \big( G_L(\hat{\lambda}_L) - G(0) \big) \\
&= \sqrt{L} \rho_0 \big( G_L(\hat{\lambda}_L) - T_L(\hat{\lambda}_L) + T_L(\hat{\lambda}_L) - G(0) \big).
\end{align*}
Lemma~\ref{lem:empiricprocforrho} shows that 
\[
\sqrt{L}\big( G_L(\hat{\lambda}_L) - T_L(\hat{\lambda}_L)  \big) = o_P(1) + X_L,
\]
where $X_L = \sqrt{L}\big( G_L(0) - T_L(0)  \big)$.  Lemma~\ref{lem:HLtoG} shows that
\[
\sqrt{L}\big( T_L(\hat{\lambda}_L) - G(0) \big) = o_P(1).
\]
It follows that $\sqrt{L} \hat{m}_L =  \rho_0 X_L + o_P(1)$.  Lemma~\ref{lem:XL} shows that the distribution of $X_L$ converges to the normal with zero mean and variance $p B_1 + d B_2$ as $L\rightarrow\infty$ where $B_1$ and $B_2$ are defined in the statement of Theorem~\ref{thm:normality}.  Thus, the distribution of $\sqrt{L} \hat{m}_L$ converges to the normal with zero mean and variance $\rho_0^2(p B_1 + d B_2)$ as required.  Because $X_L$ does not depend on $\hat{\lambda}_L$, it follows that $\covar(X_L, \sqrt{L}\hat{\lambda}_L) = 0$, and so,
\[
\covar(\sqrt{L}\hat{m}_L, \sqrt{L}\hat{\lambda}_L) \rightarrow \covar(\rho_0 X_L, \sqrt{L}\hat{\lambda}_L) = 0
\]
as $L \rightarrow \infty$.  The lemmas that we have used are proved in Section~\ref{sec:proof-asympt-normappendix} of the appendix.

\section{The Gaussian noise case}\label{sec:gaussian-noise-case}

Let the noise sequence $\{w_i\}$ be complex Gaussian with independent real and imaginary parts having zero mean and variance $\sigma^2$.  The joint density function of the real and imaginary parts is
\[
\frac{1}{2\pi\sigma^2}e^{-\frac{1}{2\sigma^2}(x^2 + y^2)}.
\]
Theorems~\ref{thm:consistency}~and~\ref{thm:normality} hold, and since the distribution of $w_1$ is circularly symmetric, the distribution of $R_1e^{j\Phi_1}$ is identical to the distribution of $1 + \frac{1}{\rho_0} w_1$.
It can be shown that
\[
g(\phi) = \frac{\cos\phi}{2\pi}e^{-\kappa} + \frac{\Psi(\sqrt{2\kappa} \cos\phi)}{\sqrt{\pi\kappa}}  e^{-\kappa\sin^2\phi} \big(2 + \kappa\cos^2\phi \big)
\]
where $\kappa = \tfrac{\rho_0^2}{2\sigma^2}$ and $\Psi(t) = \frac{1}{\sqrt{2\pi}} \int_{-\infty}^{t} e^{-x^2/2} dx$
is the cumulative density function of the standard normal.
The value of $A_1, A_2, B_1$ and $B_2$ can be efficiently computed by numerical integration using this formula.

\section{Simulations}\label{sec:simulations}

We present the results of Monte-Carlo simulations with the least squares estimator.  In all simulations the noise samples $w_1,\dots,w_L$ are independent and identically distributed circularly symmetric and Gaussian with real and imaginary parts having variance $\sigma^2$.  Under these conditions the least squares estimator is also the maximum likelihood estimator.  Simulations are run with $M=2,4,8$ (BPSK, QPSK, $8$-PSK) and with signal to noise ratio $\text{SNR} = \tfrac{\rho_0^2}{2\sigma^2}$ between \unit[-20]{dB} and \unit[20]{dB} in steps of \unit[1]{dB}.  The amplitude $\rho_0=1$ and $\theta_0$ is uniformly distributed on $[-\pi, \pi)$.  For each value of SNR, $T = 5000$ replications are performed to obtain $T$ estimates $\hat{\rho}_1, \dots, \hat{\rho}_T$ and $\hat{\theta}_1, \dots, \hat{\theta}_T$.  

\begin{figure}[p]
	\centering
		\includegraphics[width=\linewidth]{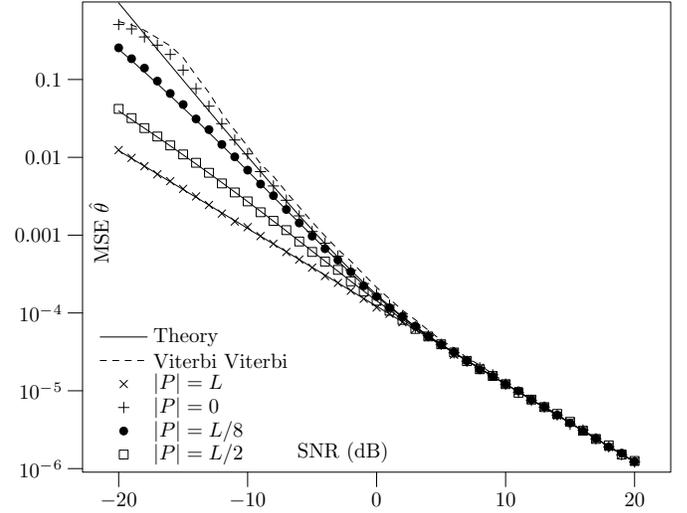}
		\caption{Phase error versus SNR for BPSK with $L=4096$.}
		\label{fig:plotphaseBPSK}
\end{figure}

\begin{figure}[p]
	\centering
		\includegraphics[width=\linewidth]{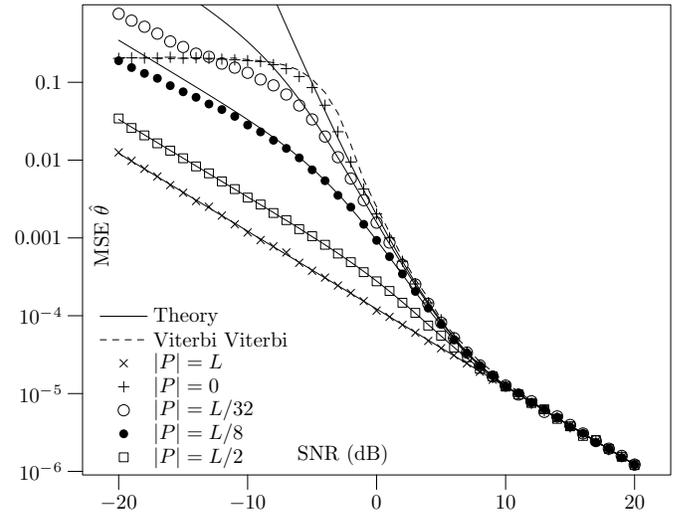}
		\caption{Phase error versus SNR for QPSK with $L=4096$.}
		\label{fig:plotphaseQPSK}
\end{figure}

\begin{figure}[p]
	\centering
		\includegraphics[width=\linewidth]{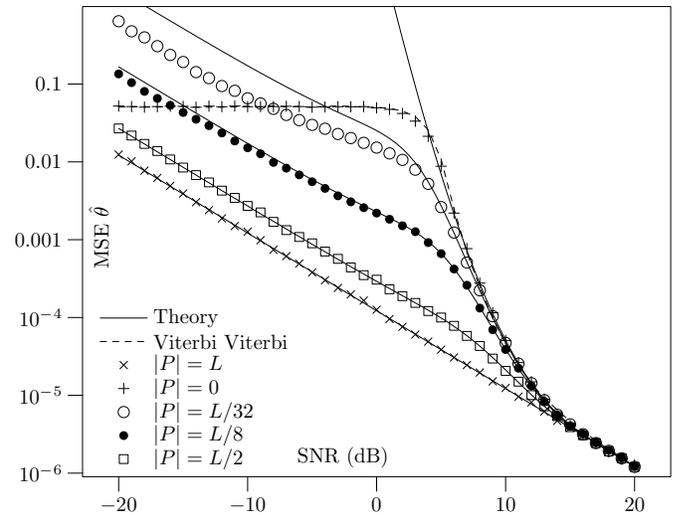}
		\caption{Phase error versus SNR for $8$-PSK with $L=4096$.}
		\label{fig:plotphase8PSK}
\end{figure}

Figures~\ref{fig:plotphaseBPSK},~\ref{fig:plotphaseQPSK}~and~\ref{fig:plotphase8PSK} show the sample mean square error (MSE) of the phase estimator when $M=2,4,8$ with $L=4096$ and for varying proportions of pilots symbols $\abs{P} = 0, \tfrac{L}{32}, \tfrac{L}{8}, \tfrac{L}{2}, L$.  When $\abs{P} \neq 0$ (i.e. coherent detection) the mean square error is computed as $\tfrac{1}{T}\sum_{i=1}^T\sfracpart{\hat{\theta}_i - \theta_0}_\pi^2$.  Otherwise, when $\abs{P}=0$ the mean square error is computed as $\tfrac{1}{T}\sum_{i=1}^T\sfracpart{\hat{\theta}_i - \theta_0}^2$ as in~\cite{McKilliam_leastsqPSKnoncoICASSP_2012}.  The dots, squares, circles and crosses are the results of Monte-Carlo simulations with the least square estimator.  The solid lines are the estimator MSEs predicted by Theorem~\ref{thm:normality}.   
The prediction is made by dividing the asymptotic covariance matrix by $L$.  The theorem accurately predicts the behaviour of the phase estimator when $L$ is sufficiently large.  As the SNR decreases the variance of the phase estimator approaches that of the uniform distribution on $[-\pi, \pi)$ when $\abs{P} \neq 0$ and the uniform distribution on $[-\tfrac{\pi}{M}, \tfrac{\pi}{M})$ when $\abs{P}=0$~\cite{McKilliam_leastsqPSKnoncoICASSP_2012}.  Theorem~\ref{thm:normality} does not model this behaviour in the sense that for any fixed $L$ there exist sufficiently small values of SNR for which Theorem~\ref{thm:normality} does not produce accurate predictions of the MSE.  As the SNR increases the variance of the estimators converge to that of the estimator where all symbols are pilots, i.e. $\abs{P} = L$.


Figures~\ref{fig:plotphaseBPSK},~\ref{fig:plotphaseQPSK}~and~\ref{fig:plotphase8PSK} also display the sample MSE of the noncoherent phase estimator of Viterbi and Viterbi~\cite{ViterbiViterbi_phase_est_1983} described by~\eqref{eq:viterbiviterbi}.  This estimator requires selection of a function $F$ that transforms the amplitude of each sample prior to the final estimation step.  
Viterbi and Viterbi propose several viable alternatives, from which we have chosen $F(x) = 1$.  The Viterbi and Viterbi estimator is only applicable in the noncoherent setting, i.e. when $\abs{P} = 0$.  The sample MSE of the least squares estimator (when $\abs{P} = 0$) and the Viterbi and Viterbi estimator is similar.  The least squares estimator appears slightly more accurate for some values of SNR.

\begin{figure}[p]
	\centering
		\includegraphics[width=\linewidth]{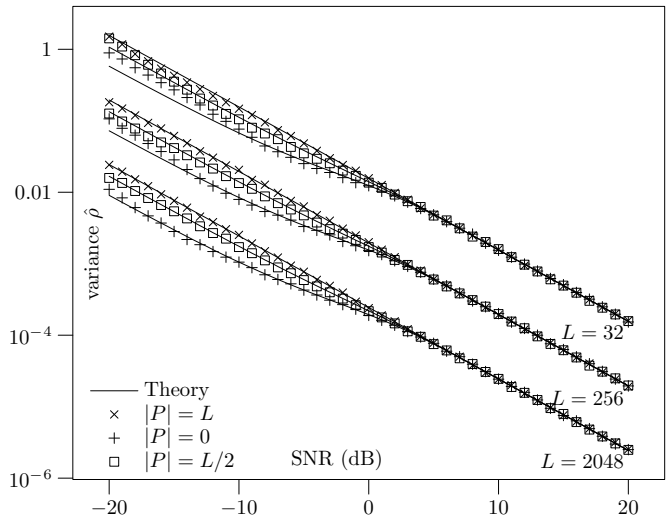}
		\caption{Unbiased amplitude error versus SNR for BPSK.}
		\label{fig:plotampBPSK}
\end{figure}

\begin{figure}[p]
	\centering
		\includegraphics[width=\linewidth]{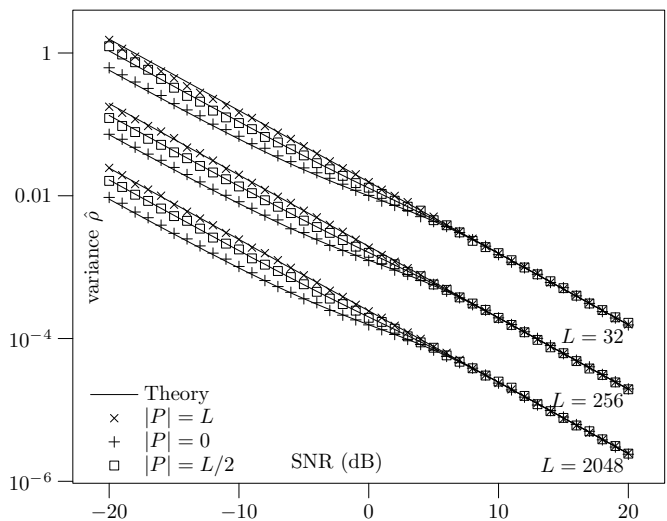}
		\caption{Unbiased amplitude error versus SNR for QPSK.}
		\label{fig:plotampQPSK}
\end{figure}

\begin{figure}[p]
	\centering
		\includegraphics[width=\linewidth]{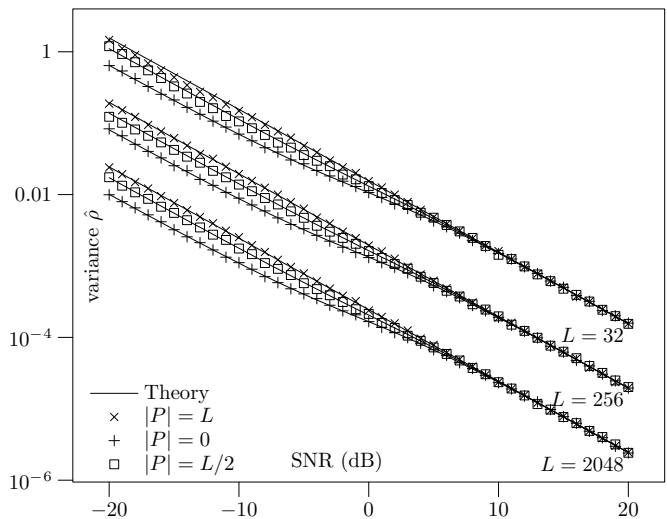}
		\caption{Unbiased amplitude error versus SNR for $8$-PSK.}
		\label{fig:plotamp8PSK}
\end{figure}

Figures~\ref{fig:plotampBPSK},~\ref{fig:plotampQPSK}~and~\ref{fig:plotamp8PSK} show the variance of the amplitude estimator when $M=2,4,8$ and with $L=32, 256, 2048$ and when the number of pilots symbols is $\abs{P} = 0, \tfrac{L}{2}, L$.  The solid lines are the variance predicted by Theorem~\ref{thm:normality}.  The dots and crosses show the results of Monte-Carlo simulations.  Each point is computed as the unbiased error $\tfrac{1}{T}\sum_{i=1}^T\big(\hat{\rho}_i - \rho_0G(0)\big)^2$.  This requires $G(0)$ to be known.  In practice $G(0)$ may not be known at the receiver, so Figures~\ref{fig:plotampBPSK},~\ref{fig:plotampQPSK}~and~\ref{fig:plotamp8PSK} serve to validate the correctness of our asymptotic theory, rather than to suggest the practical performance of the amplitude estimator.  When SNR is large $G(0)$ is close to $1$ and the bias of the amplitude estimator is small.  However, $G(0)$ grows without bound as the variance of the noise increases, so the bias is significant when SNR is small.  

\begin{figure}[tp]
	\centering
		\includegraphics[width=\linewidth]{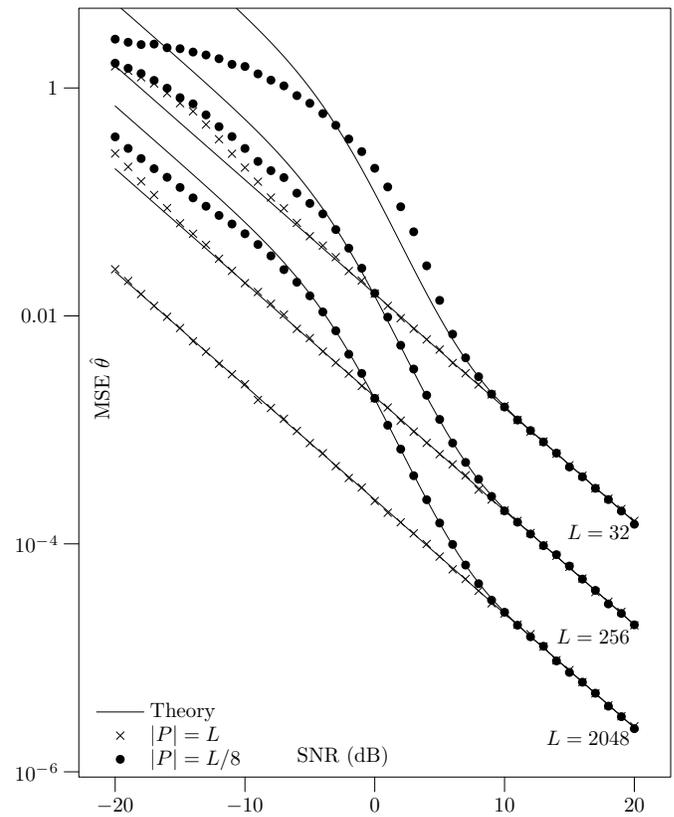}
		\caption{Phase error versus SNR for QPSK.}
		\label{fig:plotphaseQPSKmultL}
\end{figure}

Figure~\ref{fig:plotphaseQPSKmultL} shows the MSE of the phase estimator when $M=4$ and $L=32,256, 2048$ and the number of pilots is $\abs{P}=\tfrac{L}{8},L$.  The figure depicts an interesting phenomenon.  When $L=2048$ and $\abs{P} = \tfrac{L}{8} = 256$ the number of pilots symbols is the same as when $L=\abs{P} = 256$.  When the SNR is small (approximately less than \unit[0]{dB}) the least squares estimator using the $256$ pilots symbols and also the $2048-256=1792$ data symbols performs \emph{worse} than the estimator that uses only the $256$ pilots symbols.  A similar phenomenon occurs when $L=256$ and $\abs{P} = \tfrac{L}{8} = 32$.  
This behaviour suggests modifying the objective function to give the pilots symbols more importance when the SNR is low.  For example, rather than minimise~\eqref{eq:SSdefn} we could instead minimise a weighted version of it,
\[
SS_{\beta}(a, \{d_i, i \in D\}) = \sum_{i \in P} \abs{ y_i - a s_i }^2 + \beta \sum_{i \in D} \abs{ y_i - a d_i }^2,
\]
where the weight $\beta$ would be small when SNR is small and near $1$ when SNR is large.  Computing the $\hat{a}$ that minimises $SS_\beta$ can be achieved with only a minor modification to algorithm~\ref{alg:loglinear}.  Line~\ref{alg_gset} is modified to $g_i = \beta y_i e^{-j u}$ and lines~\ref{alg_hata1}~and~\ref{alg_hata2} are modified to $\hat{a} =  \frac{1}{\abs{P} + \beta\abs{D}} Y$.  For fixed $\beta$ the asymptotic properties of this weighted estimator could be derived using the techniques we have developed in Sections~\ref{sec:stat-prop-least},~\ref{sec:proof-almost-sure} and~\ref{sec:proof-asympt-norm}.  This would enable a rigorous theory for selection of $\beta$ at the receiver.  One caveat is that the receiver would require knowledge about the noise distribution in order to advantageously choose $\beta$.  We do not investigate this further here.

\section{Conclusion}

We considered least squares estimators of carrier phase and amplitude from noisy communications signals that contain both pilot signals, known to the receiver, and data signals, unknown to the receiver.  We focused on $M$-ary phase shift keying constellations.  The least squares estimator can be computed in $O(L\log L)$ operations using a modification of an algorithm due to Mackenthun~\cite{Mackenthun1994}, and is the maximum likelihood estimator in the case that the noise is additive white and Gaussian.  

We showed, under some reasonably general conditions on the distribution of the noise, that the phase estimator $\hat{\theta}$ is strongly consistent and asymptotically normally distributed.  However, the amplitude estimator $\hat{\rho}_0$ is biased, and converges to $G(0)\rho_0$.  This bias is large when the signal to noise ratio is small.  It would be interesting to investigate methods for correcting this bias.  A method for estimating $G(0)$ at the receiver appears to be required.

Monte Carlo simulations were used to assess the performance of the least squares estimator and also to validate our asymptotic theory.  Interestingly, when the SNR is small, it is counterproductive to use the data symbols to estimate the phase (Figure~\ref{fig:plotphaseQPSKmultL}).  This suggests the use of a weighted objective function, which would be an interesting topic for future research.

\small
\bibliography{bib}

\normalsize
\appendix

\subsection{Lemmas required for the proof of almost sure convergence (Theorem~\ref{thm:consistency}) } \label{sec:proof-almost-sureappendix}

\begin{lemma}\label{lem:convtoexpGlamL} 
$G(\hat{\lambda}_L) \rightarrow G(0)$ almost surely as $L \rightarrow \infty$.
\end{lemma}
\begin{IEEEproof}
Since $G(x)$ is uniquely maximised at $x=0$,
\[
0 \leq G(0) - G(\hat{\lambda}_L),
\]
and since $\hat{\lambda}_L$ is the maximiser of $G_L(x)$,
\[ 
0 \leq G_L(\hat{\lambda}_L) - G_L(0).
\]
Thus,
\begin{align*}
0 &\leq G(0) - G(\hat{\lambda}_L) \\ 
& \leq G(0) - G(\hat{\lambda}_L) + G_L(\hat{\lambda}_L) - G_L(0) \\
& \leq |G(0) - G_L(0)| + |G_L(\hat{\lambda}_L) - G(\hat{\lambda}_L)| \\
&\leq 2\sup_{\lambda \in [-\pi, \pi)} \sabs{G_L(\lambda) - G(\lambda)},
\end{align*}
and the last line converges almost surely to zero by Lemma~\ref{lem:uniflawGGL}.
\end{IEEEproof}


\begin{lemma}\label{lem:uniflawGGL} 
$\sup_{\lambda \in [-\pi, \pi)}\sabs{G_L(\lambda) - G(\lambda)} \rightarrow 0$ almost surely as $L \rightarrow \infty$.
\end{lemma}
\begin{IEEEproof}
Put $T_L(\lambda) = \expect G_L(\lambda)$ and write
\begin{align*}
&\sup_{\lambda \in [-\pi, \pi)}\sabs{G_L(\lambda) - G(\lambda)} \\
&= \sup_{\lambda \in [-\pi, \pi)}\sabs{G_L(\lambda) - T_L(\lambda) + T_L(\lambda) - G(\lambda)} \\
&\leq \sup_{\lambda \in [-\pi, \pi)}\sabs{G_L(\lambda) - T_L(\lambda)} +  \sup_{\lambda \in [-\pi, \pi)}\sabs{T_L(\lambda) - G(\lambda)}.
\end{align*}
Now,
\begin{align*}
T_L(\lambda) - G(\lambda) &= \big(\tfrac{\abs{P}}{L} - p\big)h_1(\lambda) + \big(\tfrac{\abs{D}}{L} - d\big)h_2(\lambda) \\
&= o(1) h_1(\lambda) + o(1) h_2(\lambda)
\end{align*}
Since 
\[
\sabs{h_1(\lambda)} = \sabs{\expect R_1\cos(\lambda + \Phi_1)} \leq \expect R_1,
\]
and 
\[
\sabs{h_2(\lambda)} = \sabs{\expect R_1\cos\fracpart{\lambda + \Phi_1}} \leq \expect R_1
\] 
for all $\lambda \in [-\pi, \pi)$, it follows that 
\[
\sup_{\lambda \in [-\pi, \pi)}\sabs{T_L(\lambda) - G(\lambda)} \leq o(1) \expect R_1 \rightarrow 0
\]  
as $L\rightarrow \infty$.  Lemma~\ref{lem:uniflawTGL} shows that 
\[
\sup_{\lambda \in [-\pi, \pi)}\sabs{G_L(\lambda) - T_L(\lambda)} \rightarrow 0
\]
almost surely as $L\rightarrow \infty$.
\end{IEEEproof}

\begin{lemma}\label{lem:uniflawTGL} 
Put $T_L(\lambda) = \expect G_L(\lambda)$.  Then
\[
\sup_{\lambda \in [-\pi, \pi)}\sabs{G_L(\lambda) - T_L(\lambda)} \rightarrow 0
\] 
almost surely as $L \rightarrow \infty$.
\end{lemma}
\begin{IEEEproof}
Put $D_L(\lambda) = G_L(\lambda) - T_L(\lambda)$ and let
\[
\lambda_n = \tfrac{2\pi}{N}(n-1) - \pi, \qquad n = 1, \dots, N
\]
be $N$ points uniformly spaced on the interval $[-\pi, \pi)$.  Let $L_n = [\lambda_n, \lambda_n + \tfrac{2\pi}{N})$ and observe that $L_1, \dots, L_N$ partition $[-\pi, \pi)$.  Now
\begin{align*}
\sup_{\lambda \in [-\pi, \pi)}&\sabs{D_L(\lambda)} \\
&= \sup_{n = 1,\dots,N}\sup_{\lambda \in L_n}\sabs{D_L(\lambda) - D_L(\lambda_n) + D_L(\lambda_n)} \\
&\leq U_L + V_L,
\end{align*}
where 
\[
U_L = \sup_{n = 1,\dots,N}\sabs{D_L(\lambda_n)} \;\;\; \text{and}
\]
\[
V_L = \sup_{n = 1,\dots,N}\sup_{\lambda \in L_n}\sabs{D_L(\lambda) - D_L(\lambda_n)}.
\]
Lemma~\ref{lem:supDLlambdan} shows that for any $N$ and $\epsilon > 0$, 
\[
\prob\left\{ \lim_{L\rightarrow\infty} U_L > \epsilon \right\} = 0,
\]
that is, $U_L \rightarrow 0$ almost surely as $L\rightarrow\infty$.  Lemma~\ref{lem:supsupDLlambda} shows that for any $\epsilon > 0$,
\[
\prob\left\{ \lim_{L \rightarrow \infty} V_L > \epsilon + \tfrac{4\pi }{N}\expect R_i \right\} = 0.
\] 
If we choose $N$ large enough that $4\pi\expect R_i < \epsilon N$ then
\begin{align*}
\prob& \left\{ \lim_{L\rightarrow\infty} \sup_{\lambda \in [-\pi, \pi)}\sabs{D_L(\lambda)} > 3\epsilon \right\} \\
&\leq \prob\left\{\lim_{L\rightarrow\infty}(U_L + V_L) > 3\epsilon \right\} \\
&\leq \prob\left\{\lim_{L\rightarrow\infty} (U_L + V_L) > \epsilon + \epsilon + \tfrac{4\pi }{N}\expect R_i \right\} \\
&\leq \prob\left\{\lim_{L\rightarrow\infty} U_L > \epsilon\right\} +  \prob\left\{ \lim_{L\rightarrow\infty} V_L > \epsilon + \tfrac{4\pi }{N}\expect R_i \right\} \\
&= 0.
\end{align*}
Thus $\sup_{\lambda \in [-\pi, \pi)}\sabs{D_L(\lambda)} \rightarrow 0$ almost surely as $L\rightarrow\infty$.
\end{IEEEproof}

\begin{lemma}\label{lem:supDLlambdan}
For any $N > 0$, $U_L \rightarrow 0$ almost surely as $L \rightarrow \infty$ where $U_L$ is defined in the proof of Lemma~\ref{lem:uniflawTGL}.
\end{lemma}
\begin{IEEEproof}
Put
\[
Z_i(\lambda) = \begin{cases}
R_i\cos(\lambda + \Phi_i) , & i \in P \\
R_i\cos\sfracpart{\lambda + \Phi_i} , & i \in D
\end{cases}
\]
so that
\[
D_L(\lambda) = G_L(\lambda) - T_L(\lambda) = \frac{1}{L}\sum_{i\in P \cup D} \big( Z_i(\lambda) - \expect Z_i(\lambda) \big).
\]
Now $Z_1(\lambda_n), \dots,Z_L(\lambda_n)$ are independent with finite variance (because $\expect R_i^2$ is finite), so for each $n =1, \dots, N$,
\[
\abs{D_L(\lambda_n)} = \abs{\frac{1}{L}\sum_{i\in P \cup D} \big( Z_i(\lambda_n) - \expect Z_i(\lambda_n)\big)} \rightarrow 0
\]
almost surely as $L\rightarrow\infty$ by Kolmogorov's strong law of large numbers~\cite{Billingsley1979_probability_and_measure}.  Thus
\[
U_L = \sup_{n=1,\dots,N}\sabs{D_L(\lambda_n)} \leq \sum_{n=1}^N \sabs{D_L(\lambda_n)} \rightarrow 0
\]
almost surely to zero as $L \rightarrow \infty$.
\end{IEEEproof}

\begin{lemma}\label{lem:supsupDLlambda} For any $\epsilon > 0$,
\[
\prob\left\{ \lim_{L \rightarrow \infty} V_L > \epsilon + \tfrac{4\pi}{N}\expect R_i \right\} = 0.
\]
\end{lemma}
\begin{IEEEproof}
Observe that
\begin{align*}
&\sabs{D_L(\lambda) - D_L(\lambda_n)} \\
&=  \abs{G_L(\lambda) - T_L(\lambda) - G_L(\lambda_n) + T_L(\lambda_n)} \\
&\leq  \abs{G_L(\lambda) - G_L(\lambda_n)} + \abs{\expect G_L(\lambda) - \expect G_L(\lambda_n)} \\
&\leq \abs{G_L(\lambda) - G_L(\lambda_n) } + \expect \abs{G_L(\lambda) - G_L(\lambda_n)},
\end{align*}
the last line following from Jensen's inequality.  Put
\[
C_L = \sup_{n = 1,\dots,N}\sup_{\lambda \in L_n} \abs{G_L(\lambda) - G_L(\lambda_n) },
\]
so that
\begin{align*}
V_L &= \sup_{n = 1,\dots,N}\sup_{\lambda \in L_n}\sabs{D_L(\lambda) - D_L(\lambda_n)} \\
&\leq  C_L +  \sup_{n = 1,\dots,N}\sup_{\lambda \in L_n} \expect\abs{G_L(\lambda) - G_L(\lambda_n) } \\
&\leq C_L + \expect C_L,
\end{align*}
where the last line follows because $\sup \expect\vert \dots\vert \leq \expect \sup \vert \dots \vert$.  Lemma~\ref{lem:CL} shows that $\expect C_L \leq \tfrac{2\pi}{N}\expect R_1$ and also that
\[
\prob\left\{ \lim_{L \rightarrow \infty} C_L > \epsilon + \tfrac{2\pi}{N}\expect R_1 \right\} = 0.
\]
Thus,
\begin{align*}
\prob&\left\{ \lim_{L \rightarrow \infty} V_L > \epsilon + \tfrac{4\pi}{N}\expect R_1 \right\} \\
&\leq \prob\left\{ \lim_{L \rightarrow \infty} (C_L + \expect C_L) > \epsilon + \tfrac{4\pi}{N}\expect R_1 \right\} \\
&\leq \prob\left\{ \lim_{L \rightarrow \infty} C_L > \epsilon + \tfrac{2\pi}{N}\expect R_1 \right\} = 0.
\end{align*}
\end{IEEEproof}

\begin{lemma}\label{lem:CL} The following statements hold:
\begin{enumerate}
\item $\expect C_L \leq \tfrac{2\pi}{N}\expect R_1$ for all positive integers $L$,
\item for any $\epsilon > 0$, $\prob\left\{ \lim_{L \rightarrow \infty} C_L > \epsilon + \tfrac{2\pi}{N}\expect R_1 \right\} = 0$.
\end{enumerate}
\end{lemma}
\begin{IEEEproof}
If $\lambda \in L_n$, then $\lambda = \lambda_n + \delta$ with $\delta < \tfrac{2\pi}{N}$, and from Lemma~\ref{lem:coslipshitz},
\begin{align*}
&\abs{\cos(\lambda + \Phi_i) - \cos(\lambda_n + \Phi_i)} \leq \tfrac{2\pi}{N}, \;\;\; \text{and} \\
&\abs{\cos\sfracpart{\lambda + \Phi_i} - \cos\sfracpart{\lambda_n + \Phi_i}} \leq \tfrac{2\pi}{N}.
\end{align*}
Because these results do not depend on $n$,
\[
\sup_{n=1,\dots,N}\sup_{\lambda \in L_n} \abs{Z_i(\lambda) - Z_i(\lambda_n)} \leq R_i \frac{2\pi}{N}
\]
for all $i = P \cup D$.  Also
\begin{align*}
C_L &= \sup_{n=1,\dots,N}\sup_{\lambda \in L_n}\abs{\frac{1}{L}\sum_{i \in P \cup D} Z_i(\lambda) - Z_i(\lambda_n)} \\
&\leq \frac{1}{L}\sum_{i \in P \cup D} \sup_{n=1,\dots,N}\sup_{\lambda \in L_n} \abs{Z_i(\lambda) - Z_i(\lambda_n)} \\
&\leq \frac{2\pi}{N L}\sum_{i \in P \cup D} R_i .
\end{align*}
Thus, 
\[
\expect C_L \leq \expect \frac{2\pi}{N L}\sum_{i \in P \cup D} R_i = \frac{2\pi}{N}\expect R_1
\]
and the first statement holds.  Now, 
\[
\frac{2\pi}{N L}\sum_{i \in P \cup D} R_i \rightarrow \frac{2\pi}{N}\expect R_1
\] 
almost surely as $L \rightarrow\infty$ by the strong law of large numbers, and so, for any $\epsilon > 0$,
\begin{align*}
\prob&\left\{ \lim_{L \rightarrow \infty} C_L > \epsilon + \tfrac{2\pi}{N}\expect R_1 \right\} \\
&\leq \prob\left\{ \lim_{L \rightarrow \infty} \frac{2\pi}{N L}\sum_{i \in P \cup D} R_i > \epsilon + \tfrac{2\pi}{N}\expect R_1 \right\} = 0.
\end{align*}
\end{IEEEproof}

\begin{lemma}\label{lem:coslipshitz}
Let $x$ and $\delta$ be real numbers.  Then
\begin{align*}
&\abs{\cos(x + \delta) - \cos(x)} \leq \abs{\delta}, \;\;\; \text{and} \\
&\abs{\cos\fracpart{x + \delta} - \cos\fracpart{x}} \leq \abs{\delta}.
\end{align*}
\end{lemma}
\begin{IEEEproof}
Both $\cos(x)$ and $\cos\fracpart{x}$ are Lipschitz continuous functions from $\reals$ to $\reals$ with constant $K=1$.  That is, for any $x$ and $y$ in $\reals$,
\begin{align*}
&\abs{\cos(y) - \cos(x)} \leq K\abs{x-y} = \abs{x-y}, \;\;\; \text{and} \\
&\abs{\cos\fracpart{y} - \cos\fracpart{x}} \leq K\abs{x-y} = \abs{x-y}.
\end{align*}
 The lemma follows by putting $y = x + \delta$.
\end{IEEEproof}

\begin{lemma}\label{lem:GLtoG0}
$G_L(\hat{\lambda}_L) \rightarrow G(0)$ almost surely as $L \rightarrow \infty$.
\end{lemma}
\begin{IEEEproof}
By the triangle inequality,
\[
\sabs{G_L(\hat{\lambda}_L) - G(0)} \leq \sabs{G_L(\hat{\lambda}_L) - G(\hat{\lambda}_L)} + \sabs{G(\hat{\lambda}_L) - G(0)}.
\]
Now $\sabs{G_L(\hat{\lambda}_L) - G(\hat{\lambda}_L)} \rightarrow 0$ as $L \rightarrow \infty$ as a result of Lemma~\ref{lem:uniflawGGL}, and $\sabs{G(\hat{\lambda}_L) - G(0)} \rightarrow 0$ almost surely as $L \rightarrow \infty$ because $G$ is continuous and $\hat{\lambda}_L \rightarrow 0$ almost surely as $L \rightarrow \infty$.
\end{IEEEproof}

\subsection{Lemmas required for the proof of asymptotic normality (Theorem~\ref{thm:normality}) } \label{sec:proof-asympt-normappendix}

\begin{lemma}~\label{lem:diffatlambdaL}
The derivative of $G_L$ exists, and is equal to zero, at $\hat{\lambda}_L$.  That is,
\[
G_L^\prime(\hat{\lambda}_L) = \frac{d G_L}{d \lambda}(\hat{\lambda}_L) = 0.
\]
\end{lemma}
\begin{IEEEproof}
Observe that 
\[
G_L(\lambda) = \frac{1}{L}\sum_{i \in P} R_i \cos(\lambda + \Phi_i) + \frac{1}{L} \sum_{i \in D} R_i \cos\sfracpart{\lambda + \Phi_i}
\] 
is differentiable everywhere except when $\fracpart{\lambda + \Phi_i} = -\tfrac{\pi}{M}$ for any $i \in D$ with $R_i > 0$.  Let $q_i$ be the smallest number from the interval $[-\tfrac{\pi}{M}, 0]$ such that
\[
 \sin(q_i) > -\frac{\rho_0 \sabs{\eta}^2 R_i}{4 L \hat{\rho} \sin(\pi/M)}
\]
where $\eta = e^{-j2\pi/M} - 1$.  Observe that $q_i < 0$ when $R_i > 0$.  Lemma~\ref{lem:fracpartlambdahatnotpi} shows that
\[
\sabs{\sfracpart{\hat{\lambda}_L + \Phi_i}} \leq \frac{\pi}{M} + q_i
\]
for all $i \in D$.  Thus, $\sfracpart{\hat{\lambda}_L + \Phi_i} \neq -\tfrac{\pi}{M}$ for $i \in D$ such that $R_i > 0$ and therefore $G_L$ is differentiable at $\hat{\lambda}_L$.  That $G_L^\prime(\hat{\lambda}_L) = 0$ follows since $\hat{\lambda}_L$ is a maximiser of $G_L$.
\end{IEEEproof}

\begin{lemma}\label{lem:fracpartlambdahatnotpi} Let $q_i$ be defined as in Lemma~\ref{lem:diffatlambdaL}.  Then $\sabs{\sfracpart{\hat{\lambda}_L + \Phi_i}} \leq \frac{\pi}{M} + q_i$ for all $i \in D$.
\end{lemma}
\begin{IEEEproof}
Recall that $\{\hat{d}_i = \hat{d}_i(\hat{\theta}), i \in D\}$ defined in \eqref{eq:hatdfinxtheta} are the minimisers of the function 
\[
SS(\{d_i, i \in D\}) = A - \frac{1}{L}\sabs{Y(\{d_i, i \in D\})}^2,
\]
defined in~\eqref{eq:SSdatasymbols}. The proof now proceeds by contradiction.  Assume that 
\begin{equation}\label{eq:fraclambassumption}
\sfracpart{\hat{\lambda}_L + \Phi_k} > \frac{\pi}{M} + q_k
\end{equation}
for some $k \in D$.  Recalling the notation $e_k$ defined in Section~\ref{sec:least-squar-estim}, put $r_i = \hat{d}_i e_k$.  We will show that 
\[
SS(\{r_i, i \in D\}) < SS(\{\hat{d}_i, i \in D\}),
\]
violating the fact that $\{\hat{d}_i, i \in D\}$ are minimisers of $SS$.  First observe that,
\begin{align*}
Y(\{ r_i, i \in D\}) = \sum_{i \in P} y_ip_i^* + \sum_{i \in D} y_i\hat{r}_i^* = \hat{Y} + \eta y_k\hat{d}_k^*,
\end{align*}
where $\eta = e^{-j2\pi/M} - 1$ and $\hat{Y} = Y(\{ \hat{d}_i, i \in D\})$.  Now,
\begin{align*}
SS&(\{r_i, i \in D\}) \\
&= A - \frac{1}{L} \sabs{ Y(\{ r_i, i \in D\}) }^2 \\
&= A - \frac{1}{L} \sabs{ \hat{Y} + \eta y_k\hat{d}_k^* }^2 \\
&= A - \frac{1}{L}\sabs{\hat{Y}}^2 - \frac{2}{L}\Re\left(\eta \hat{Y}^* y_k \hat{d}_k^* \right) -  \frac{1}{L}\sabs{ \eta y_k}^2\\
&= SS(\{\hat{d}_i, i \in D\}) - C,
\end{align*}
where 
\[
C = \frac{2}{L}\Re\left(\eta \hat{Y}^* y_k \hat{d}_k^* \right) +  \frac{1}{L}\sabs{ \eta y_k }^2.
\]
Now $\frac{1}{L}\hat{Y} = \hat{a} = \hat{\rho} e^{j\hat{\theta}}$ from~\eqref{eq:hata} and using~\eqref{eq:yethetadhat},
\[
\frac{1}{L} \hat{Y}^* y_k \hat{d}_k^* = \hat{\rho} \rho_0 R_k e^{j\sfracpart{\hat{\lambda}_L + \Phi_k}},
\]
so that
\begin{equation}\label{eq:Ddefn}
C = 2 \hat{\rho} \rho_0 R_k \Re\left( \eta e^{j\sfracpart{\hat{\lambda}_L + \Phi_k}}\right) + \frac{1}{L}\sabs{\eta}^2\rho_0^2 R_k^2.
\end{equation}
Let $v = \sfracpart{\hat{\lambda}_L + \Phi_k} - \tfrac{\pi}{M}$ so that
\begin{align*}
\eta e^{j\sfracpart{\hat{\lambda}_L + \Phi_k}} &= (e^{-j2\pi/M} - 1) e^{j \pi/M} e^{jv} \\
&= (e^{-j\pi/M} - e^{j\pi/M}) e^{jv} \\
&= -2 j \sin(\tfrac{\pi}{M}) e^{j v },
\end{align*}
and
\[
\Re\left( \eta e^{j\sfracpart{\hat{\lambda}_L + \Phi_k}}\right) = 2 \sin(\tfrac{\pi}{M}) \sin(v).
\]
Because we assumed~\eqref{eq:fraclambassumption}, it follows that $0 > v > q_k$ and, from the definition of $q_k$,
\[
 -\frac{\rho_0 \sabs{\eta}^2 R_k}{4 L \hat{\rho} \sin(\pi/M)} <  \sin(v) < 0.
\]
Substituting this into~\eqref{eq:Ddefn} gives $C > 0$, but then 
\[
SS(\{r_i, i \in D\}) < SS(\{\hat{d}_i, i \in D\}),
\] 
violating the fact that $\{\hat{d}_i, i \in D\}$ are minimisers of $SS$.  So~\eqref{eq:fraclambassumption} is false.

To show that $\sfracpart{\hat{\lambda}_L + \Phi_k} \geq -\frac{\pi}{M} - q_k$ we use contradiction again.  Assume that $\sfracpart{\hat{\lambda}_L + \Phi_k} < -\frac{\pi}{M} - q_k$.  Recalling the notation $e_k^*$ defined in Section~\ref{sec:least-squar-estim}, put $r_i = \hat{d}_i e_k^*$.  Now an analogous argument can be used to show that $SS(\{r_i, i \in D\}) < SS(\{\hat{d}_i, i \in D\})$ again.
\end{IEEEproof}

\begin{lemma}\label{lem:Qconv}
Let $Q_L(\lambda) = \expect R_L(\lambda) - \expect R_L(0)$ where the function $R_L$ is defined in~\eqref{eq:RLdef}.  We have
\[ 
\sqrt{L} Q_L(\hat{\lambda}_L) = \sqrt{L} \hat{\lambda}_L\big( p + Hd  + o_P(1) \big)
\]
where $p, d$ and $H$ are defined in the statements of Theorems~\ref{thm:consistency}~and~\ref{thm:normality}.
\end{lemma}
\begin{IEEEproof}
We have
\[
Q_L(\lambda) = \expect R_L(\lambda) - \expect R_L(0) = \tfrac{\sabs{P}}{L} k_1(\lambda) + \tfrac{\sabs{D}}{L} k_2(\lambda)
\]
where
\begin{align}
k_1(\lambda) &= \expect R_1\big( \sin(\lambda + \Phi_1) - \sin(\Phi_1) \big), \;\;\; \text{and} \nonumber\\
k_2(\lambda) &= \expect R_1\big( \sin\fracpart{\lambda + \Phi_1} - \sin\fracpart{\Phi_1} \big). \label{eq:k2def}
\end{align}
Lemma~\ref{lem:q1k1parts} shows that $k_1(\hat{\lambda}_L) = \hat{\lambda}_L\big(1 + o_P(1) \big)$ and Lemma~\ref{lem:q2k2parts} shows that $k_2(\hat{\lambda}_L) = \hat{\lambda}_L \big( H + o_P(1) \big)$ and so
\begin{align*}
Q_L(\hat{\lambda}_L) &=  \tfrac{\sabs{P}}{L} \hat{\lambda}_L \big( 1  + o_P(1)\big)  + \tfrac{\sabs{D}}{L} \hat{\lambda}_L ( H + o_P(1) ) \\
&= \hat{\lambda}_L \big( \tfrac{\sabs{P}}{L}  + \tfrac{\sabs{D}}{L} H + o_P(1) \big) \\
 &= \hat{\lambda}_L \big( p  + d H + o_P(1) \big),
 \end{align*}
since $\frac{\sabs{P}}{L} \rightarrow p$ and $\frac{\sabs{D}}{L} \rightarrow d$ as $L \rightarrow \infty$.  The lemma follows by multiplying both sides of the 
above equation by $\sqrt{L}$.
\end{IEEEproof}

\begin{lemma}\label{lem:q1k1parts}
Put
\[
q_1(\lambda) + j k_1(\lambda) = \expect\big[ R_1 e^{j(\lambda + \Phi_1)} - R_1 e^{j\Phi_1}  \big].
\]
We have $q_1(\lambda) = \hat{\lambda}_L o_P(1)$ and $k_1(\hat{\lambda}_L) = \hat{\lambda}_L\big(1 + o_P(1) \big)$.
\end{lemma}
\begin{IEEEproof}
We have
\[
q_1(\lambda) + j k_1(\lambda) = (e^{j\lambda} - 1) \expect R_1e^{j\Phi_1} = e^{j\lambda} - 1
\]
since the mean of $R_1e^{j\Phi_1}$ is 1.  By a first order expansion about $\lambda = 0$ we obtain
\[
q_1(\lambda) + j k_1(\lambda) = \lambda\big( j + O(\lambda) \big)
\]
Since $\hat{\lambda}_L$ converges almost surely to zero as $L\rightarrow\infty$ it follows that $O(\hat{\lambda}_L) = o_P(1)$.  Thus
\[
q_1(\hat{\lambda}_L) + j k_1(\hat{\lambda}_L) = \hat{\lambda}_L\big( j + o_P(1) \big)
\]
and the lemma follow by taking real and imaginary parts.
\end{IEEEproof}

\begin{lemma}\label{lem:q2k2parts}
Put
\[
q_2(\lambda) + j k_2(\lambda) = \expect\big[ R_1 e^{j\fracpart{\lambda + \Phi_1}} - R_1 e^{j\fracpart{\Phi_1}}  \big].
\]
We have $q_2(\hat{\lambda}_L) = \hat{\lambda}_L o_P(1)$ and $k_2(\hat{\lambda}_L) = \hat{\lambda}_L \big( H + o_P(1) \big)$ where $H$ is defined in the statement of Theorem~\ref{thm:normality}.
\end{lemma}
\begin{IEEEproof}
Because $\hat{\lambda}_L \rightarrow 0$ almost surely as $L\rightarrow\infty$, it is only the behaviour of $q_2(\lambda)$ and $k_2(\lambda)$ around $\lambda = 0$ that is relevant.  We will examine $q_2(\lambda)$ and $k_2(\lambda)$ for $0 \leq \lambda < \tfrac{\pi}{M}$.  An analogous argument follows when $-\tfrac{\pi}{M} < \lambda < 0$.  To keep our notation clean put 
\[
\psi_k = \frac{2\pi}{M}k + \frac{\pi}{M}
\] 
with $k \in \ints$.  When $\Phi_1 \in [\psi_{k-1}, \psi_k - \lambda)$,
\begin{align*}
e^{j\fracpart{\lambda + \Phi_1}} - e^{j\fracpart{\Phi_1}} &=  e^{j\big(\lambda + \Phi_1 - \tfrac{2\pi}{M}k\big)} - e^{j\big( \Phi_1- \tfrac{2\pi}{M}k\big)} \\
&= \big(e^{j\lambda} - 1 \big) e^{j\big(\Phi_1- \tfrac{2\pi}{M}k\big)} \\
&= \big(e^{j\lambda} - 1 \big) e^{j\fracpart{\Phi_1}},
\end{align*}
and when $\Phi_1 \in [\psi_{k} - \lambda, \psi_k)$,
\begin{align*}
e^{j\fracpart{\lambda + \Phi_1}} - &e^{j\fracpart{\Phi_1}} \\
&=  e^{j\big(\lambda + \Phi_1 - \tfrac{2\pi}{M}k - \tfrac{2\pi}{M}\big)} - e^{j\big( \Phi_1- \tfrac{2\pi}{M}k\big)} \\
&= \big(e^{j\big(\lambda - \tfrac{2\pi}{M}\big)} - 1 \big) e^{j\fracpart{\Phi_1}} \\
&= \big(e^{j\lambda} - 1 \big) e^{j\fracpart{\Phi_1}}  + e^{j\lambda}\big( e^{-j\tfrac{2\pi}{M}} - 1 \big)e^{j\fracpart{\Phi_1}}.
\end{align*}
Thus, when $\Phi_1 \in [\psi_{k-1}, \psi_k)$,
\begin{align*}
e^{j\fracpart{\lambda + \Phi_1}} - e^{j\fracpart{\Phi_1}} &= \big(e^{j\lambda} - 1 \big) e^{j\fracpart{\Phi_1}} \\
&\hspace{0.5cm} + e^{j\lambda}\big( e^{-j\tfrac{2\pi}{M}} - 1 \big)e^{j\fracpart{\Phi_1}} \chi_k(\Phi_1,\lambda)
\end{align*}
where 
\[
\chi_k(\Phi_1,\lambda) = \begin{cases}
1, & \Phi_1 \in [\psi_{k} - \lambda, \psi_k)  \\
0, & \text{otherwise}.
\end{cases}
\]
Now
\begin{align}
q_2(\lambda) + j k_2(\lambda) &= \expect\big[ R_1 e^{j\fracpart{\lambda + \Phi_1}} - R_1 e^{j\fracpart{\Phi_1}}  \big] \nonumber \\
&= \big(e^{j\lambda} - 1 \big) h_2(0)  + e^{j\lambda} B(\lambda) \label{eq:q2k2withB}
\end{align}
since $\expect R_1 e^{j\fracpart{\Phi_1}} = h_2(0) + \expect R_1 \sin\fracpart{\Phi_1} = h_2(0)$ as a result of Lemma~\ref{lem:expectImRfracpart} and where
\begin{equation}\label{eq:Blambdadefn}
B(\lambda) = \big( e^{-j\tfrac{2\pi}{M}} - 1 \big)\expect R_1 e^{j\fracpart{\Phi_1}} \chi(\Phi_1,\lambda)
\end{equation}
and $\chi(\Phi_1, \lambda) = \sum_{k \in \ints}\chi_k(\Phi_1, \lambda)$. 

Now $e^{j\hat{\lambda}_L} = 1 + o_P(1)$ and $e^{j\hat{\lambda}_L} - 1 = \hat{\lambda}_L \big( j  + o_P(1) \big)$ by the argument in Lemma~\ref{lem:q1k1parts}.  Also
\[
B(\hat{\lambda}_L) = -\hat{\lambda}_L\left(2j\sin(\tfrac{\pi}{M}) \sum_{k=0}^{M-1} g(\psi_k) + o_P(1)\right) 
\]
by Lemma~\ref{lem:Blambdaconv}.  Combining these results into~\eqref{eq:q2k2withB} we obtain
\begin{align*}
q_2(\lambda) + &jk_2(\hat{\lambda}_L) \\
&= \hat{\lambda}_L \left( jh_2(0) - 2j\sin(\tfrac{\pi}{M}) \sum_{k=0}^{M-1} g(\psi_k) + o_P(1) \right) \\
&= \hat{\lambda}_L\big( jH + o_P(1) \big)
\end{align*}
and the lemma follows by taking real and imaginary parts.
\end{IEEEproof}

\begin{lemma}\label{lem:Blambdaconv}
With $B(\lambda)$ defined in~\eqref{eq:Blambdadefn} we have 
\[
B(\hat{\lambda}_L) = -\hat{\lambda}_L\left(2j\sin(\tfrac{\pi}{M}) \sum_{k=0}^{M-1} g(\psi_k) + o_P(1)\right) 
\]
\end{lemma}
\begin{IEEEproof}
Put $A(\lambda) = \expect R_1 e^{j\fracpart{\Phi_1}} \chi(\Phi_1,\lambda)$.  Recalling that $f(r,\phi)$ is the joint pdf of $R_1$ and $\Phi_1$ we have
\begin{align*}
 A(\lambda)  &= \int_{0}^{2\pi}\int_{0}^{\infty} r f(r,\phi) e^{j\fracpart{\phi}}\chi(\phi,\lambda) dr d\phi \\
&= \sum_{k\in\ints}\int_{0}^{2\pi} g(\phi) e^{j\fracpart{\phi}} \chi_k(\phi,\lambda) d\phi \\
&= \sum_{k=0}^{M-1}\int_{\psi_k-\lambda}^{\psi_k} g(\phi)  e^{j\fracpart{\phi}} d\phi,
\end{align*}
the last line because the $\chi_k(\phi,\lambda)$ terms inside the integral are zero for all $\phi \in [0,2\pi]$ when $k \notin \{0,\dots,M-1\}$.  Observe that $\fracpart{\phi} \rightarrow \tfrac{\pi}{M}$ as $\phi$ approaches $\psi_k$ from below.  Because $g(\psi_k)$ is continuous at $\psi_k$ for each $k = 0, \dots, M-1$ (by assumption in Theorem~\ref{thm:normality}) we have 
\[
\frac{1}{\lambda} A(\lambda) \rightarrow \sum_{k=0}^{M-1} g(\psi_k) e^{j\tfrac{\pi}{M}}
\]
as $\lambda$ approaches zero from above.  We are only interested in the limit from above because we are working under the assumption that $0 \leq \lambda < \frac{\pi}{M}$ (see the proof of Lemma~\ref{lem:q2k2parts}).  The analogous argument when $-\tfrac{\pi}{M} \leq \lambda < 0$ would involve limits as $\lambda$ approaches zero from below.  Thus
\[
A(\hat{\lambda}_L) = \hat{\lambda}_L \left(  e^{j\tfrac{\pi}{M}} \sum_{k=0}^{M-1} g(\psi_k) + o_P(1) \right)
\]
and the lemma follows since $\big( e^{-j\tfrac{2\pi}{M}} - 1 \big) e^{j\frac{\pi}{M}} = -2j\sin(\tfrac{\pi}{M})$ and $B(\lambda) = \big( e^{-j\tfrac{2\pi}{M}} - 1 \big)A(\lambda)$.
\end{IEEEproof}

\begin{lemma}\label{lem:expectImRfracpart}
$\expect R_1 \sin\fracpart{\Phi_1} = 0$.
\end{lemma}
\begin{IEEEproof}
Recalling that $f(r,\phi)$ is the joint pdf of $R_1$ and $\Phi_1$ we have
\begin{align*}
\expect R_1 \sin\fracpart{\Phi_1} &= \int_{0}^{2\pi} \int_{0}^\infty r \sin\fracpart{\phi} f(r,\phi) dr d\phi \\
&= \int_{-\pi}^{\pi} \sin\fracpart{\phi} g(\phi) d\phi.
\end{align*}
The proof is immediate since $g(\phi)$ is even and $\sin\fracpart{\phi}$ is odd.
\end{IEEEproof}

\begin{lemma}\label{lem:empiricprocc} Let $Q_L(\lambda) = \expect R_L(\lambda) - \expect R_L(0)$ where the function $R_L$ is defined in~\eqref{eq:RLdef}.  Then,
\[
\sqrt{L}\big( R_L(\hat{\lambda}_L) - Q_L(\hat{\lambda}_L) \big) = \sqrt{L} R_L(0) + o_P(1).
\]
\end{lemma}
\begin{IEEEproof}
Write
\[
\sqrt{L}\big( R_L(\hat{\lambda}_L) - Q_L(\hat{\lambda}_L) \big) = W_L(\hat{\lambda}_L) + \sqrt{L} R_L(0)
\]
where
\begin{equation}\label{eq:WLdef}
W_L(\lambda) = \sqrt{L}\big( R_L(\lambda) - Q_L(\lambda) - R_L(0) \big)
\end{equation}
is what is called an \emph{empirical process} indexed by $\lambda$~\cite{Pollard_asymp_empi_proc_1989,Pollard_new_ways_clts_1986,van2009empirical,Pollard_conv_stat_proc_1984}.  Techniques from this literature can be used to show that
for any $\delta > 0$ and $\nu > 0$, there exists $\epsilon > 0$ such that
\[
\Pr\left\{ \sup_{\sabs{\lambda}<\epsilon}\sabs{W_L(\lambda)} > \delta  \right\} < \nu
\]
for all positive integers $L$.  This type of result is typically called \emph{tightness} or \emph{asymptotic continuity}~\cite{Pollard_asymp_empi_proc_1989,van2009empirical,Billingsley1999_convergence_of_probability_measures}.  We omit the proof which follows in a straightforward, but lengthy manner using an argument called \emph{symmetrisation} followed by an argument called \emph{chaining}~\cite{Pollard_asymp_empi_proc_1989,van2009empirical}.

Since $\hat{\lambda}_L$ converges almost surely to zero, it follows that for any $\epsilon > 0$,
\[
\lim_{L\rightarrow\infty}\prob\left\{ \sabs{\hat{\lambda}_L} \geq \epsilon \right\} = 0
\] 
and therefore, for any $\nu > 0$, $\prob\{ \sabs{\hat{\lambda}_L} \geq \epsilon \} < \nu$ for all sufficiently large $L$.  Now
\begin{align*}
  \prob&\left\{\sabs{ W_L(\hat{\lambda}_L) } > \delta \right\} \\
&= \prob\left\{ \sabs{W_L(\hat{\lambda}_L)} > \delta \;\text{and} \; \sabs{\hat{\lambda}_L} < \epsilon \right\} \\
&\hspace{0.7cm} + \prob\left\{ \sabs{W_L(\hat{\lambda}_L)} > \delta  \;\text{and} \; \sabs{\hat{\lambda}_L} \geq \epsilon \right\} \\
&\leq \prob\left\{  \sup_{\sabs{\lambda} < \epsilon} \sabs{ W_L(\lambda) } > \delta \right\} + \prob\left\{ \sabs{\hat{\lambda}_L} \geq \epsilon \right\} \\
&\leq 2\nu,
\end{align*}
for all sufficiently large $L$.  Since $\nu$ and $\delta$ can be chosen arbitrarily small, it follows that $W_L(\hat{\lambda}_L)$ converges in probability to zero as $N\rightarrow\infty$.
\end{IEEEproof}

\begin{lemma}\label{lem:convdistGLdash}
The distribution of $\sqrt{L}R_L(0)$ converges to the normal with zero mean and variance $pA_1 + dA_2$ as $L\rightarrow\infty$.
\end{lemma}
\begin{IEEEproof}
Observe that $\sqrt{L} R_L(0) = C_L + D_L$ where
\[
C_L = \frac{1}{\sqrt{L}} \sum_{i \in P} R_i \sin(\Phi_i), \qquad D_L = \frac{1}{\sqrt{L}} \sum_{i \in D} R_i \sin\sfracpart{\Phi_i}.
\]
From the standard central limit theorem the distribution of $C_L$ converges to the normal with mean $\sqrt{p}\expect R_1 \sin(\Phi_1) = 0$ as a result of~\eqref{eq:expectImpartRphi}, and variance
\[
p A_1 = p \expect R_1^2 \sin^2(\Phi_1).
\]
Similarly, the distribution of $D_L$ converges to the normal with mean $\sqrt{d}\expect R_1 \sin\fracpart{\Phi_1} = 0$ as a result of Lemma~\ref{lem:expectImRfracpart}, and variance
\[
d A_2 = d \expect R_1^2 \sin^2\fracpart{\Phi_1}.
\]
The lemma holds because $C_L$ and $D_L$ are independent. 
\end{IEEEproof}

\begin{lemma}\label{lem:empiricprocforrho} Let $T_L(\lambda) = \expect G_L(\lambda)$.  We have
\[
\sqrt{L}\big( G_L(\hat{\lambda}_L) - T_L(\hat{\lambda}_L) \big) =  X_L + o_P(1),
\]
where $X_L = \sqrt{L} \big( G_L(0) - T_L(0) \big)$.
\end{lemma}
\begin{IEEEproof}
Write
\[
\sqrt{L}\big( G_L(\hat{\lambda}_L) - T_L(\hat{\lambda}_L) \big) = Y_L(\hat{\lambda}_L) + X_L
\]
where
\begin{equation}\label{eq:YLdef}
Y_L(\lambda) = \sqrt{L}\big( G_L(\lambda) - T_L(\lambda) \big) - X_L.
\end{equation}
is an empirical process indexed by $\lambda$, similar to $W_L$ from~\eqref{eq:WLdef}.  As with $W_L$ results from the literature on empirical processes can be used to show that  
for any $\delta > 0$ and $\nu > 0$, there exists $\epsilon > 0$ such that
\[
\Pr\left\{ \sup_{\sabs{\lambda}<\epsilon}\sabs{Y_L(\lambda)} > \delta  \right\} < \nu.
\]
The proof now follows by an argument analogous to that in Lemma~\ref{lem:empiricprocc}.
\end{IEEEproof}

\begin{lemma}\label{lem:HLtoG} $\sqrt{L}\big( T_L(\hat{\lambda}_L) - G(0) \big) = o_P(1)$.
\end{lemma}
\begin{IEEEproof}
The argument is similar to that used in Lemma~\ref{lem:Qconv}.  First observe that
\begin{align*}
T_L(\lambda) &= \frac{\sabs{P}}{L}\expect R_1\cos(\lambda + \Phi_1) + \frac{\sabs{D}}{L} \expect R_1\cos\fracpart{\lambda + \Phi_1} \\
&=  \big( p + o(L^{-1/2}) \big) \expect R_1\cos(\lambda + \Phi_1) \\
&\hspace{2cm} + \big( d + o(L^{-1/2}) \big) \expect R_1\cos\fracpart{\lambda + \Phi_1},
\end{align*}
and because $\tfrac{\abs{P}}{L} = p + o(L^{-1/2})$ and $\tfrac{\abs{D}}{L} = d + o(L^{-1/2})$ (by assumption in Theorem~\ref{thm:normality}), we have
\begin{align*}
\sqrt{L}\big( T_L(\lambda) - G(0) \big) = \sqrt{L}p q_1(\lambda) + \sqrt{L} d q_2(\lambda) + o(1),
\end{align*}
where
\begin{align}
q_1(\lambda) &= \expect R_1\big( \cos(\lambda + \Phi_1) - \cos(\Phi_1) \big) \;\;\; \text{and} \nonumber \\
q_2(\lambda) &= \expect R_1\big( \cos\fracpart{\lambda + \Phi_1} - \cos\fracpart{\Phi_1} \big). \label{eq:q2def}
\end{align}
Lemma~\ref{lem:q1k1parts} shows that $q_1(\hat{\lambda}_L) = \hat{\lambda}_Lo_P(1)$ and Lemma~\ref{lem:q2k2parts} shows that $q_2(\hat{\lambda}_L) = \hat{\lambda}_L o_P(1)$ and so
\[
\sqrt{L}\big( T_L(\lambda) - G(0) \big) =\sqrt{L}\hat{\lambda}_L o_P(1) + o(1).
\]
The lemma follows since $\sqrt{L}\hat{\lambda}_L$ converges in distribution.
\end{IEEEproof}

\begin{lemma}\label{lem:XL} 
The distribution of 
\[
X_L = \sqrt{L} \big( G_L(0) - T_L(0) \big) = \sqrt{L} \big( G_L(0) - \expect G_L(0) \big)
\] 
converges, as $L \rightarrow\infty$,  to the normal with zero mean and covariance $pB_1 + d B_2.$
\end{lemma}
\begin{IEEEproof}
Observe that $X_L = C_L^\prime + D_L^\prime$ where
\[
C_L^\prime = \frac{1}{\sqrt{L}} \sum_{i \in P} \big( R_i \cos(\Phi_i) - h_1(0) \big),
\]
\[
D_L^\prime = \frac{1}{\sqrt{L}} \sum_{i \in D} (R_i \cos\fracpart{\Phi_i} - h_2(0) ),
\]
where $h_1$ and $h_2$ are defined in the statement of Theorem~\ref{thm:consistency}.  From the standard central limit theorem the distribution of $C_L^\prime$ converges to the normal with zero mean and variance
\[
p B_1 = p \expect R_1^2 \cos^2(\Phi_1) - p h_1^2(0) =  p \expect R_1^2 \cos^2(\Phi_1) - p 
\]
since $h_1(0) = 1$.  Similarly the distribution of $D_L^\prime$ converges to the normal with zero mean and variance
\[
d B_2 = d \expect R_1^2 \cos^2\fracpart{\Phi_1} - d h_2^2(0).
\]
The lemma follows since $C_L^\prime$ and $D_L^\prime$ are independent.
\end{IEEEproof}

\end{document}